\documentclass[12pt]{article} 

\usepackage{epsfig} 
\usepackage{amsbsy}  
 
\newlength{\largfig}
\def\nul{\nu_\ell}

\def\ds#1{#1\kern-1ex\hbox{/}} 
\def\sl#1{#1\kern-1ex\hbox{/}} 
\def\dsh{h\kern-1.2ex /}

\def\beq{\begin{equation}} 
\def\eeq{\end{equation}} 
\def\eq{\beq\eeq} 
\def\beqn{\begin{eqnarray}} 
\def\eeqn{\end{eqnarray}} 
 
\def\lq{\left[} 
\def\rq{\right]} 
\def\rg{\right\}} 
\def\lg{\left\{} 
\def\({\left(} 
\def\){\right)} 
 
\def\ba{\begin{eqnarray}} 
\def\ea{\end{eqnarray}} 
\def\bq{\begin{equation}} 
\def\eq{\end{equation}} 
 
\def\gsim{\mathrel{\raisebox{-.6ex}{$\stackrel{\textstyle>}{\sim}$}}} 
\def\sla#1{\ifmmode% 
\setbox0=\hbox{$#1$}% 
\setbox1=\hbox to\wd0{\hss$/$\hss}\else% 
\setbox0=\hbox{#1}% 
\setbox1=\hbox to\wd0{\hss/\hss}\fi% 
#1\hskip-\wd0\box1 }

\def\MB{{\cal M}_B}

\def\I{{\cal I}} 
\def\I{{\boldsymbol{I}}}

\def\asb{{}\ifmmode \bar{\alpha}_s \else $\bar{\alpha}_s$\fi} 
\def \as   {\ifmmode \alpha_s \else $\alpha_s$ \fi} 
 
\def\xd#1{\tilde D_{#1}(k_2,q_2,q_1)}

\def\epsbkb{\epsilon_2\cdot k_2}
\def\epsbka{\epsilon_2\cdot k_1}
\def\epsbqa{\epsilon_2\cdot q_1}
\def\epsakb{\epsilon_1\cdot k_2}
\def\epsaka{\epsilon_1\cdot k_1}
\def\epsaqb{\epsilon_1\cdot q_2}
\def\epsaepsb{\epsilon_1\cdot \epsilon_2}

\def\so3#1{\,{\rm S}_{1,\,3}\left(#1 \right)} 
\def\st2#1{\,{\rm S}_{2,\,2}\left(#1 \right)} 

\def\Re{\mathop{\rm Re}}

\newskip\humongous \humongous=0pt plus 1000pt minus 1000pt

\newif\ifdtup

\catcode`@=12 
 
\catcode`\@=11 
 
\@addtoreset{equation}{section} 

\def\theequation{\thesection.\arabic{equation}} 
 
\def\@normalsize{\@setsize\normalsize{15pt}\xiipt\@xiipt 
\abovedisplayskip 14pt plus3pt minus3pt% 
\belowdisplayskip \abovedisplayskip 
\abovedisplayshortskip \z@ plus3pt% 
\belowdisplayshortskip 7pt plus3.5pt minus0pt} 
 
\def\small{\@setsize\small{13.6pt}\xipt\@xipt 
\abovedisplayskip 13pt plus3pt minus3pt% 
\belowdisplayskip \abovedisplayskip 
\abovedisplayshortskip \z@ plus3pt% 
\belowdisplayshortskip 7pt plus3.5pt minus0pt 
\def\@listi{\parsep 4.5pt plus 2pt minus 1pt 
     \itemsep \parsep 
     \topsep 9pt plus 3pt minus 3pt}} 
 
\@twosidetrue

\catcode`\@=11  
\def\section{\@startsection{section}{1}{\z@}{3.5ex plus 1ex minus 
   .2ex}{2.3ex plus .2ex}{\large\bf}}

\def\thesection{\arabic{section}} 
\def\thesubsection{\arabic{section}.\arabic{subsection}} 
\def\thesubsubsection{\arabic{section}.\arabic{subsection}.\arabic{subsubsection}} 
 
\def\appendix{\setcounter{section}{0} 
 \def\thesection{\Alph{section}} 
 \def\theequation{\Alph{section}.\arabic{equation}} 
\def\thesubsection{\Alph{section}.\arabic{subsection}} 
\def\thesubsubsection{\Alph{section}.\arabic{subsection}.\arabic{subsubsection}} 
 
\def\section{\@startsection{section}{1}{\z@}{3.5ex plus 1ex minus 
   .2ex}{2.3ex plus .2ex}{\large\bf}} 
}

\newcommand{\ccaption}[2]{ 
  \begin{center} 
    \parbox{0.85\textwidth}{ 
      \caption[#1]{\small\it {#2}}} 
  \end{center}    } 
 
\def \ep{\epsilon} 
\def \eps{\epsilon} 
\def \to   {\mbox{$\rightarrow$}}

\def\ord#1{{\cal O}\(#1\)} 
 
\newcount\minutes 
\newcount\scratch 
 
\def\timestamp{% 
\scratch=\time 
\divide\scratch by 60 
\edef\hours{\the\scratch} 
\multiply\scratch by 60 
\minutes=\time 
\advance\minutes by -\scratch 
---$\,$\hours:\null 
\ifnum\minutes< 10 0\fi 
\the\minutes}

\begin{document} 
\begin{titlepage} 
\nopagebreak 
{\flushright{ 
        \begin{minipage}{5cm} 
	 DCPT /03/110\\
	 IPPP /03/55\\
         MADPH 03-1348  \\ 	 
        {\tt hep-ph/0310156}\hfill \\ 
        \end{minipage}        } 
 
} 
\vfill 
\begin{center} 
{\LARGE \bf \sc 
 \baselineskip 0.9cm 
QCD corrections to electroweak\\
$\ell\nul jj$ and $\ell^+\ell^- jj$ 
production

} 
\vskip 0.5cm  
{\large   
Carlo Oleari$^1$ and Dieter Zeppenfeld$^2$ 
}   
\vskip .2cm  
{$^1$ {\it Department of Physics, University of Durham,
South Road, Durham DH1 3LE, UK}}\\ 
{$^2$ {\it Department of Physics, University of Wisconsin, Madison, WI 
53706, USA }}\\

\vskip 
1.3cm     
\end{center} 
 
\nopagebreak 
%\vfill 
%\vskip 3cm 
\begin{abstract}
The production of $W$ 
%$(\to \ell \nul)$ 
or $Z$ 
%$(\to \ell^+\ell^-)$ 
bosons in association with two jets is an important background to the Higgs
boson search in vector-boson fusion at the LHC.  The purely electroweak
component of this background is dominated by vector-boson fusion, which
exhibits kinematic distributions very similar to the Higgs boson signal. 
We consider the next-to-leading order QCD corrections to the electroweak
production of $\ell \nul jj$ and $\ell^+\ell^- jj$ events at the LHC, 
within typical vector-boson fusion cuts.
We show that the QCD corrections are modest, increasing the total cross
sections by about 10\%.  Remaining scale uncertainties are below 2\%. A
fully-flexible next-to-leading order partonic Monte Carlo program allows to
demonstrate these features for cross sections within typical
vector-boson-fusion acceptance cuts. Modest corrections are also found for
distributions.
\end{abstract} 
\vfill 
%\today \timestamp \hfill 
\vfill 

% PACS: 14.80.Bn
\end{titlepage} 
\newpage

\section{Introduction} 
Vector-boson fusion (VBF) processes have emerged as a particularly
interesting class of scattering events from which one hopes to gain insight
into the dynamics of electroweak symmetry breaking. The most prominent
example is Higgs boson production via VBF, that is, the process $qq\,\to\,
qqH$, which can be viewed as quark scattering via $t$-channel exchange of a
weak boson, with the Higgs boson radiated off the $W$ or $Z$
propagator. Alternatively, one may view this process as two weak bosons
fusing to form the Higgs boson.  The kinematic characteristics of this
process are very distinctive: two jets, in the forward and backward region of
rapidity, with the Higgs boson decay products in the central region of the
detector.  This characteristic signature greatly helps to distinguish these
$Hjj$ events from backgrounds.  Higgs boson production via VBF has been
studied intensively as a tool for Higgs boson discovery~\cite{ATLAS,CMS} and
the measurement of Higgs boson couplings~\cite{Zeppenfeld:2000td} in $pp$
collisions at the CERN Large Hadron Collider (LHC).

Analogous to Higgs boson production via VBF, the electroweak production of a
$W$ or $Z$ plus two jets, with the requirement that the weak boson is
centrally produced and that the two jets are well separated in rapidity, will
proceed with sizable cross section at the LHC\footnote{Another source of
$Wjj$ or $Zjj$ events are QCD processes at order $\alpha_s^2\alpha$,
sometimes called QCD $Vjj$ production. Within typical VBF cuts, cross
sections for these QCD processes are only somewhat larger than those for
electroweak production~\cite{Rainwater:1999gg}.  One thus needs to calculate
NLO QCD corrections for both sources independently, and as a function of
phase space. For the QCD processes this was done in
Ref.~\cite{Campbell:2002tg}.}.
The decay 
leptons in $W \,\to\,\ell\nul$ and $Z\,\to\,\ell^+\ell^-$ lead to 
the final states $\ell \nul jj$ and $\ell^+\ell^-jj$ ($\ell=e,\mu,\tau$). 
These processes have already been 
considered in the literature at leading order (LO). To name but a few
examples, they have been studied in the investigation of rapidity gaps at
hadron colliders~\cite{Chehime:1992ub,Rainwater:1996ud, Khoze:2002fa}, as a
probe of anomalous triple-gauge-boson couplings~\cite{Baur:1993fv} or as a
background to Higgs boson searches in
VBF~\cite{wbfhtautau,wbfhtoww,Eboli:2000ze}.  In this last case, 
the $\ell\nul jj$ final state with an unidentified charged lepton,
or $\nul\bar\nul jj$ events from $Z\to \nul\bar\nul$ decay, form a background 
to invisible Higgs boson decay (see e.g.\ Ref.~\cite{Eboli:2000ze}). 
$\tau^+\tau^-jj$ events are a background to the decay
$H\,\to\,\tau^+\tau^-$~\cite{wbfhtautau}, and also to $H\,\to\, W^+ W^-$ when
the $W$'s and the $\tau$'s decay leptonically~\cite{wbfhtoww}.
In these examples, off-shell corrections to $Z\,\to\,\tau^+\tau^-$ decay
need to be included, since a Higgs boson mass in the range 
$114~{\rm GeV} < m_H < 200~{\rm GeV}$, well above the $Z$ peak, 
is favored by electroweak data~\cite{Charlton:2001am}. 

While a LO analysis is perfectly adequate for exploratory investigation,
precision measurements at the LHC require comparison with cross-section
predictions which include higher-order QCD corrections. A poignant example is
the extraction of Higgs boson couplings, where expected accuracies of the
order of 10\%, or even better~\cite{Zeppenfeld:2000td}, clearly require
knowledge of the next-to-leading order (NLO) QCD corrections.
In addition, one would like to exploit $W$ and $Z$ production, in VBF
configurations,  as calibration processes for Higgs
boson production via VBF, namely as a tool to understand the tagging of
forward jets or the distribution and veto of additional central jets in VBF
(see e.g.\ Ref.~\cite{Rainwater:1996ud,Khoze:2002fa}).  
In fact, these processes share the same color structure: two colored quarks
are scattered via the exchange of a colorless boson
in the $t$-channel. The pattern of soft gluon radiation 
%in the collinear and at large scattering angle 
is then the same. Understanding the gap-survival
probability in the known case of $W$ and $Z$ production can give insight on
the gap survival for the case of Higgs boson production.
The precision needed for Higgs boson studies and for the understanding of
radiation patterns then requires the knowledge of NLO QCD corrections for
$Wjj$ and $Zjj$ production as well.

%In a recent paper~\cite{Figy:2003nv}, we presented the calculation of the NLO
%QCD corrections for the $Hjj$ cross section in the form of a fully-flexible 
%parton-level Monte Carlo program. We here extend this work and describe
%the calculation and first results for such corrections to $Wjj$ and $Zjj$ 
%production via VBF. To be precise, we consider the electroweak
%processes $pp\,\to\, \ell^\pm \nu jjX$ and $pp\,\to\, \ell^+\ell^-jjX$ at
%NLO,  
%i.e., we include off-shell corrections to the decaying $W$ or $Z$ bosons, 
%as depicted in Fig.~\ref{fig:feynBorn}(e,f). Nevertheless, we will generically
%call these processes ``EW $Vjj$ production'' in the following, ignoring
%the inclusion of off-shell effects in our nomenclature. 
%

The NLO QCD corrections to the total $Hjj$ cross section from VBF has been
known for many years~\cite{Han:1992hr}. In a recent paper~\cite{Figy:2003nv},
we presented the calculation of these corrections in the form of a
fully-flexible parton-level Monte Carlo program which allows the
determination of NLO corrections to arbitrary (infrared-safe) distributions.
Here, we extend this work and describe the calculation and first results for
such corrections to $Wjj$ and $Zjj$ production in VBF configurations.  To be
precise, since the decaying weak bosons are spin-one particles, in order to
retain all the possible angular correlations between the final state
particles, we consider the electroweak processes $pp\,\to\, \ell^\pm \nul
jjX$ and $pp\,\to\, \ell^ +\ell^-jjX$ at NLO.
%%%%%%%%%%%%%%%%%%%%%%%%%%%%%%%%%%%%%%%5
%%%%%%%%%%%%%%%%%%%%%%%%%%%%%%%%%%%%%%%5
\begin{figure}[!htb] 
%\vspace*{-0.3in}
\centerline{ 
\epsfig{figure=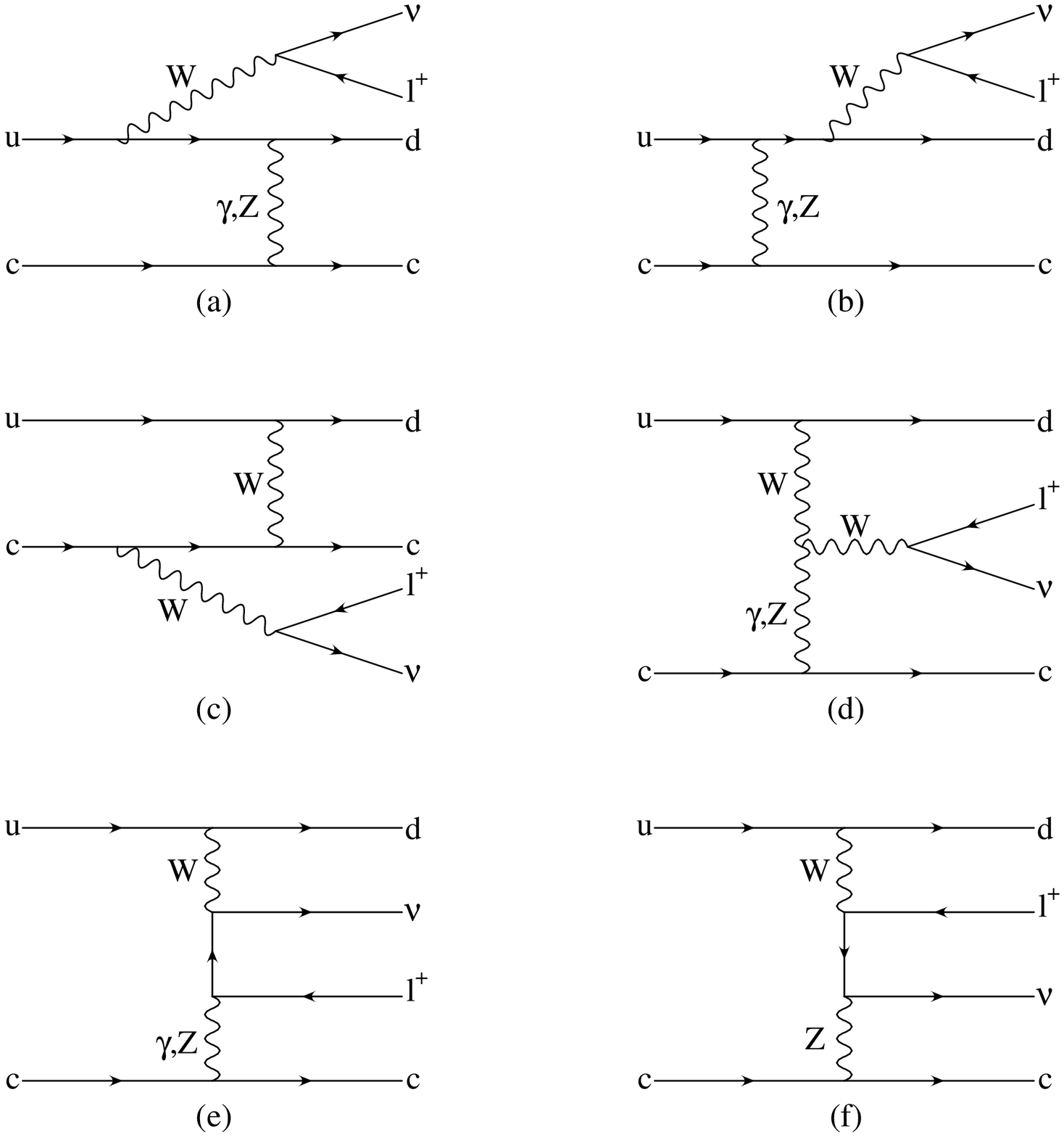,width=0.79\textwidth,clip=} 
} 
\ccaption{} 
{\label{fig:feynBorn} 
Feynman graphs contributing to the process 
$uc\,\to\, dc\ell^+\nul$ at tree level. For the generic VBF process discussed 
in this paper, seven 
Feynman-graph topologies contribute at tree level: the six topologies shown 
plus an additional bremsstrahlung graph, with the vector boson 
emitted off the final-state charm quark [mirror image of graph (b)]. 
}
\end{figure} 
%%%%%%%%%%%%%%%%%%%%%%%%%%%%%%%%%%%%%%%5
%%%%%%%%%%%%%%%%%%%%%%%%%%%%%%%%%%%%%%%5
At LO, Feynman graphs for one such process, $uc\,\to\, dcW^+,W^+\,\to\,
\ell^+\nul$, are shown in Fig.~\ref{fig:feynBorn}.  Using the terminology
introduced in~\cite{Boudjema:1996qg}, we consider bremsstrahlung (a, b, c),
fusion~(d) and multiperipheral (e, f) diagrams.  We neglect diagrams
corresponding to conversion, abelian and non-abelian annihilation, since
these $q\bar q$ annihilation contributions are negligible when we impose 
VBF cuts, as explained in detail in Sec.~\ref{sec:approximations}.

In the following, in order to use a shorthand notation, we will call processes 
such as the one depicted in Fig.~\ref{fig:feynBorn} ``EW $Vjj$ production'', or VBF
production of $W/Z$ plus two jets, since we consider these processes with the
kinematic cuts typical for the selection of VBF (see Sec.~\ref{sec:pheno}). It should be
understood that, in spite of this notation, multiperipheral diagrams 
like~(e) and~(f) are included, even though they cannot be represented as the
production of a weak boson, followed by its decay into two leptons.

The structure of the paper is as follows: 
in Sec.~\ref{sec:calculation}, we outline the calculation of the tree-level
diagrams,  of real-emission contributions and of the virtual corrections.
We dedicate Sec.~\ref{sec:virtual} to the  discussion of the virtual
contributions, with some of the analytical details relegated to
Appendix~\ref{sec:appendix}.
A list of checks which we have performed on our calculation concludes
Sec.~\ref{sec:calculation}.
Additional features of our Monte Carlo program, like the gauge invariant
handling of finite $W$ and $Z$ widths, the inclusion of anomalous $WW\gamma$ 
and $WWZ$ couplings, the approximations with regard to crossed
diagrams in the presence of identical quark flavors, the singularities for
incoming photons and the choice of parameters, will be discussed in
Sec.~\ref{sec:MC}.  We then use this Monte Carlo program to present first
results for EW $Vjj$ production at the LHC. Of particular concern is the
scale dependence of the NLO results, which provides an estimate for the
residual theoretical error of our cross-section calculations. We discuss
the scale dependence and the size of
the radiative corrections for various distributions in
Sec.~\ref{sec:pheno}. Conclusions are given in Sec.~\ref{sec:summary}.

\section{Elements of the calculation}
\label{sec:calculation}
The structure of the three processes under consideration, $pp\,\to\,\ell^+\nul
jjX$, $pp\,\to\,\ell^-\bar\nul jjX$ and $pp\,\to\, \ell^+\ell^-jjX$, is very
similar. A discussion of any single one of them is sufficient to clarify our
procedures for all, and we use $W^+$ production, i.e., the calculation of the
$pp\,\to\,\ell^+\nul jjX$ cross section, for this purpose. {\em Mutatis
mutandis}, all the considerations apply to the other processes too.

\subsection{Approximations and general framework}
\label{sec:approximations}
At tree level, the topological structure of the generic subprocesses
contributing to EW $Wjj$ production is depicted in Fig.~\ref{fig:feynBorn}.
Two additional classes of diagrams appear in case of identical
quark flavors on two of the fermion lines:
\begin{itemize}
%\item[-] Diagrams where an incoming quark and antiquark annihilate to form a
%virtual $W/Z$ boson, and another vector boson subsequently decays into a
%final-state quark-antiquark pair. These diagrams correspond to vector-boson
%pair production.
%%%%%%%%%%%%%%%%%%%%%%%%%%%%%%%%%%%%%%%%%%%%%%%%%%%%%%%%%%%
\item[-] diagrams where both the two virtual vector bosons are time-like. They
  correspond to diagrams called conversion, abelian and
  non-abelian annihilation in Ref.~\cite{Boudjema:1996qg}, and contain 
  vector-boson pair production with subsequent decay of one of the weak
  bosons to a pair of jets. {\it Pars pro toto}, we call this class
 vector-boson pair production in the following.

\item[-] diagrams obtained by interchange of identical initial- or 
final-state (anti)quarks, such as in the $uu\,\to\, du\ell^+\nul$ or
$d u \, \to \, d d \ell^+ \nul$ subprocesses.
\end{itemize}
These additional diagrams are obtained from the ones shown in
Fig.~\ref{fig:feynBorn} by crossing. In our calculation, we have neglected  
contributions from vector-boson pair production completely. In addition,
any interference effects of the second class with the graphs of 
Fig.~\ref{fig:feynBorn} are neglected. 
This is justified because, in the phase-space 
region where VBF can be observed experimentally, with widely-separated quark 
jets of very large invariant mass, the neglected terms are strongly 
suppressed by large momentum transfer in one or more
weak-boson propagators. Color suppression further reduces any interference
terms.
We have checked with MadEvent~\cite{madgraph} that, at LO, the diagrams that
we have not considered and interference effects contribute less than 0.3\% 
to our final results in e.g.\ Fig.~\ref{fig:scale_depW}.
% My result is 1060 pb
% MadEvent is  1057 pb
Since we expect QCD corrections to the neglected terms to be modest,
%same order of magnitude as the corrections which we have found in this paper, 
the above approximations are fully justified within the accuracy 
of our NLO calculation.

Fermion masses are set to zero throughout, because observation of either
leptons or (light) quarks in a hadron-collider environment requires large 
transverse momenta and hence sizable scattering angles and relativistic 
energies. 
For the $t$-channel processes
which we include, we have used a diagonal form (equal to the identity matrix)
for the Cabibbo-Kobayashi-Maskawa matrix, $V_{CKM}$. This approximation is
not a limitation of our calculation. As long as no final-state quark flavor
is tagged (no $c$ tagging is done, for example), the sum over all flavors,
using the exact $V_{CKM}$, is equivalent to our results, due to the unitarity
of the $V_{CKM}$ matrix.

\subsection{Tree-level diagrams and real corrections}
\begin{figure}[htb] 
%\vspace*{-0.1in}
\centerline{ 
\epsfig{figure=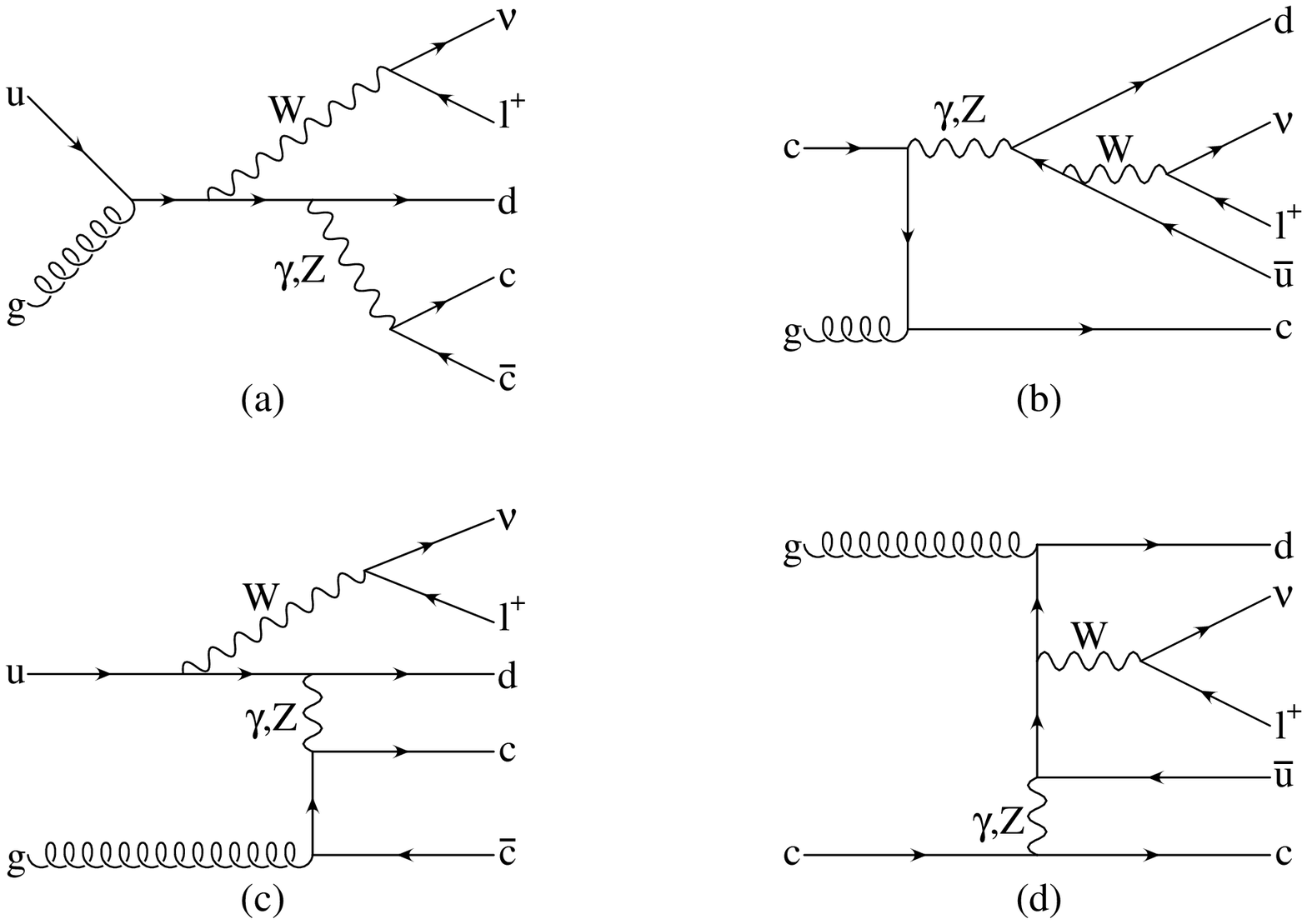,width=0.8\textwidth,clip=} \ \  
} 
\ccaption{} 
{\label{fig:feyndrop} 
Examples of Feynman amplitudes with an initial gluon. Graphs like~(a) and~(b),
with the gluon coupled to the initial quark line, correspond to 
vector-boson pair production and are eliminated. The two gauge-invariant
subsets of  
graphs like~(c) and~(d), with the gluon coupled to the final-state quark 
pair, contain all $g\to q\bar q$ splitting contributions and are included
in our calculation.
}
\end{figure}

For the $Wjj$ Born amplitude, we need to add the contributions from the 10 
Feynman graphs shown in Fig.~\ref{fig:feynBorn} ($Z$ and $\gamma$ propagators 
counted as different diagrams), and sum cross sections of all subprocesses
producing $W^+$ plus two jets. The same is
true for $W^-$ production. For the case of $Zjj$ production, amplitudes
which correspond to neutral-current exchange (no change of quark flavors)
receive contributions from 24 Feynman graphs at tree level.
To obtain the real-emission diagrams, with a final-state gluon, one needs to
attach the gluon to the quark lines in all possible ways. For the diagrams in
Fig.~\ref{fig:feynBorn}, this gives rise to 45 real-emission graphs.
112 different Feynman graphs contribute to real-emission corrections
to $Zjj$ production via neutral-current exchange.

The contributions with an initial-state gluon are obtained by crossing the
previous diagrams, promoting the final-state gluon as incoming parton, and an
initial-state (anti)quark as final-state particle.  
We again remove all diagrams where two time-like, final-state vector bosons
appear such as $g u \,\to \,\ell^+ \nul d Z^*$,
with $Z^* \,\to\, c \bar c$. Such diagrams, for consistency, must be
removed since we have not considered the corresponding Born
contributions. Figure~\ref{fig:feyndrop} clarifies this issue: we drop all
initial-gluon contributions in which the gluon couples to the fermion line of
the initial quark or antiquark.  In fact, these diagrams are strongly
suppressed when VBF cuts (see Sec.~\ref{sec:pheno}) are applied to the
final-state jets.

Our Monte Carlo program computes all amplitudes numerically, using the 
formalism of Ref.~\cite{HZ}. The Born amplitudes for $W$ and $Z$ production 
are taken from Ref.~\cite{Chehime:1992ub}. The real-emission amplitudes for
$Z$ production were first given in Ref.~\cite{Rainwater:1996ud}. The 
corresponding amplitudes for $W$ production were partially programed at the
time. We have finalized and tested them for the present application.

\subsection{Virtual corrections}
\label{sec:virtual}

At NLO, we have to deal with soft and collinear singularities in the virtual
and real-emission contributions.
Our calculation uses the subtraction method of Catani and Seymour~\cite{CS}
to cancel the soft and collinear divergences between virtual and
real-emission diagrams. Since these divergences only depend on the color 
structure of the external partons, the subtraction terms encountered for EW
$Vjj$ production are identical in form to those found for Higgs boson
production in VBF. Thus, we can use the results described in
Ref.~\cite{Figy:2003nv} for the case at hand. 
The main difference is that the finite parts of the virtual corrections are
more complicated than for $Hjj$ production (where only vertex corrections were
present).

The QCD corrections to EW $Vjj$ production appear as two gauge-invariant
subsets, 
corresponding to corrections to the upper and lower fermion lines in
Fig.~\ref{fig:feynBorn}.  Due to
the color singlet nature of the exchanged electroweak bosons, any 
interference terms
between subamplitudes with gluons attached to both the upper and the lower
quark lines vanish identically at order $\alpha_s$. Hence, it is sufficient to
consider radiative corrections to a single quark line only, which we here 
take as the upper one. Corrections to the lower fermion line are an exact copy.

In computing the virtual corrections, we have used the dimensional reduction
scheme~\cite{DR_citation}: we have performed the
Passarino-Veltman reduction of the tensor  integrals in $d=4-2\ep$
dimensions, while the algebra of the Dirac gamma matrices, of the external
momenta and of the polarization vectors has been performed in $d=4$ dimensions.

We split the virtual corrections into two  classes: the virtual
corrections along a quark line with only one weak boson attached and the
virtual corrections along a quark line with two weak bosons attached.

{\bf I.} The virtual NLO QCD contribution to any tree level Feynman
subamplitude ${\cal M}_B^{(i)}$ which has a single electroweak boson $V$
(of momentum $q$) attached to the upper fermion line,
\beq 
q(k_1) \,\to\, q(k_2) + V(q) \;,
\eeq
appears in the form of a vertex correction, which is factorisable in terms of
the original Born subamplitude
\bq
\label{eq:vertexvirt} 
{\cal M}_V^{(i)} =
{\cal M}_B^{(i)} \frac{\alpha_s(\mu_R)}{4\pi} C_F
\(\frac{4\pi\mu_R^2}{Q^2}\)^\epsilon \Gamma(1+\epsilon)
\lq-\frac{2}{\epsilon^2}-\frac{3}{\epsilon}+c_{\rm virt} 
%+ i\pi\(-\frac{2}{\ep} -3 \)
 +\ord{\ep}\rq \,.
\eq 
Here $\mu_R$ is the renormalization scale, and the boson virtuality 
$Q^2 = -(k_1-k_2)^2 = - q^2$ is the only relevant scale in the process,
since the quarks are assumed to be massless, $k_1^2 = k_2^2=0$.
In dimensional reduction, the finite contribution is given by $c_{\rm
virt}=\pi^2/3-7$ ($c_{\rm virt}=\pi^2/3-8$ in conventional dimensional
regularization).
%In the following, we will neglect the imaginary part, since it drops out when
%the interference with the Born is performed, and the complex conjugate
%contribution is added.

\begin{figure}[t] 
\centerline{ 
\epsfig{figure=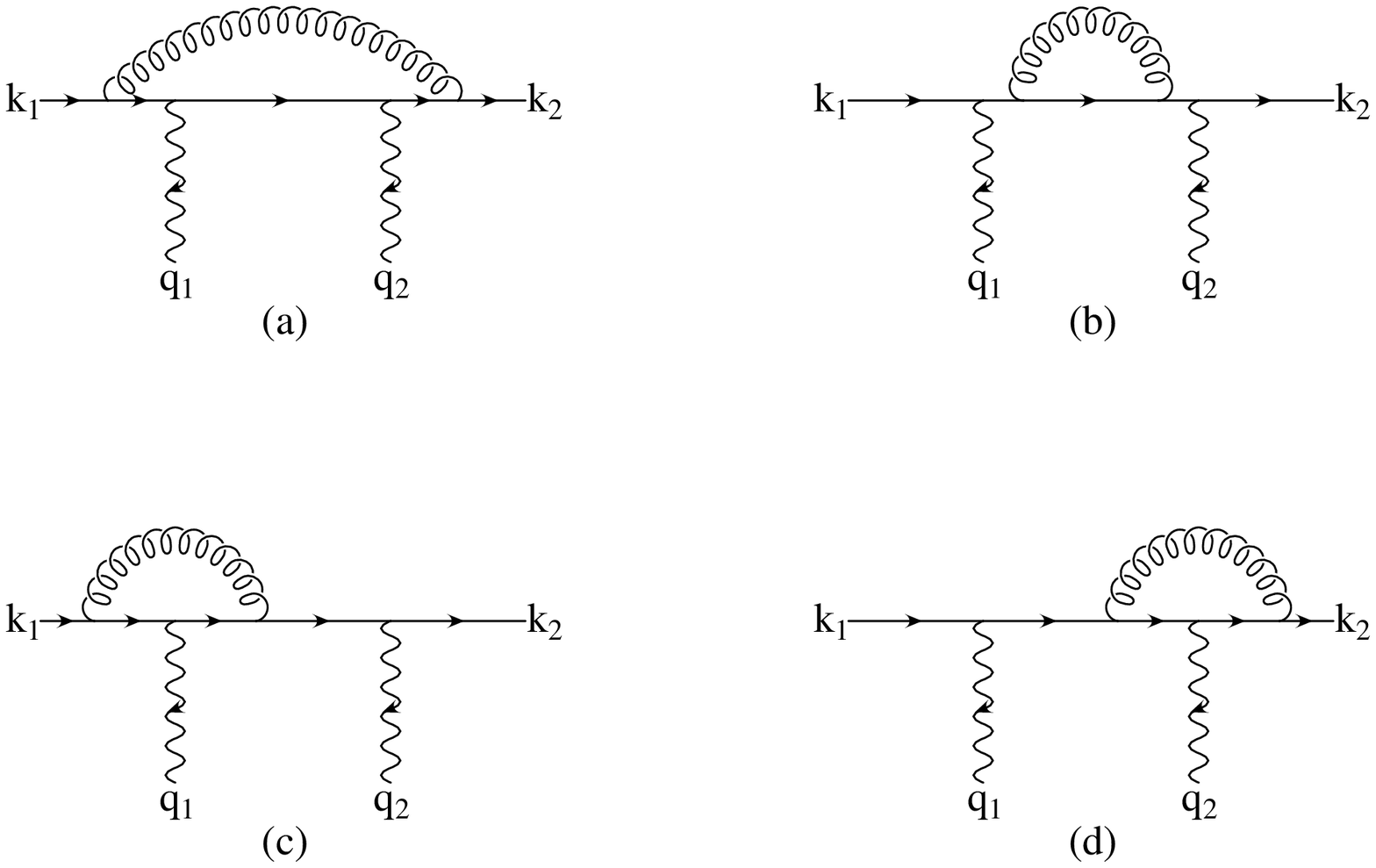,width=0.8\textwidth,clip=} \ \  
} 
\ccaption{} 
{\label{fig:boxline} 
Virtual corrections for a fermion line with two 
attached electroweak bosons, $V_1(q_1)$ and $V_2(q_2)$. The finite part
of the sum of these graphs defines the reduced amplitude 
$\tilde{\cal M}_\tau(q_1,q_2)$ of Eq.~(\ref{eq:boxlinefig}).
}
\end{figure}

{\bf II.} The second class of diagrams are the virtual QCD corrections to the 
Feynman graphs where two electroweak
bosons $V_1$ and $V_2$ (of outgoing momenta $q_1$ and $q_2$) are attached to
the same fermion line (see, for example, the upper quark line in
Fig.~\ref{fig:feynBorn} (a, b)).
It suffices to consider one of the two possible permutations of $V_1$ and
$V_2$, as depicted in Fig.~\ref{fig:boxline}.
The kinematics is given by
\bq
q(k_1)\,\to \, q(k_2)+V_1(q_1)+V_2(q_2)\,,
\eq
where $k_1^2=k_2^2=0$ and momentum conservation reads $k_1=k_2+q_1+q_2$.
In the following, it is convenient to use the Mandelstam variables
for a $2\,\to\, 2$ process which we take as $q\bar q \, \to \, V_1 V_2$.
We then define
\bq
\label{eq:kinematics}
s = (k_1-k_2)^2 = (q_1+q_2)^2\;,\quad
t = (k_1-q_1)^2 = (k_2+q_2)^2\;,\quad
u = (k_1-q_2)^2 = (k_2+q_1)^2\;.
\eq
In order to use the same notation as in Eq.~(\ref{eq:vertexvirt}), we define
$Q^2=2k_1\cdot k_2\equiv-s$.

The two electroweak bosons are always virtual in our calculation, i.e., the
effective polarization vectors $\eps_1(q_1)$ and $\eps_2(q_2)$ actually
correspond to fermion currents (the charm-quark current and the
leptonic-decay currents in the Feynman graphs of
Fig.~\ref{fig:feynBorn} (a, b)).   
Since fermion masses are neglected, current conservation implies transversity
of the effective polarization vectors: $\eps_1\cdot q_1 =\eps_2\cdot q_2 =0$. 
The expressions that we give in Appendix~\ref{sec:appendix} exploit this
relationship.
Our numerical code permits to switch on the missing $\eps_1\cdot q_1$ and
$\eps_2\cdot q_2$ terms, allowing us to test gauge invariance.
Due to the trivial color structure of the corresponding tree-level
diagram, the divergent part (soft and collinear singularities) of the sum of
the four diagrams in Fig.~\ref{fig:boxline} is a multiple of this Born
subamplitude, just like for the vertex corrections,
\ba
\label{eq:boxlinefig} 
{\cal M}_{\rm boxline}^{(i)} &=&
{\cal M}_B^{(i)} \frac{\alpha_s(\mu_R)}{4\pi} C_F
\(\frac{4\pi\mu_R^2}{Q^2}\)^\epsilon \Gamma(1+\epsilon)
\lq-\frac{2}{\epsilon^2}-\frac{3}{\epsilon}+c_{\rm virt} \rq \nonumber \\
&+&\frac{\alpha_s(\mu_R)}{4\pi} C_F
\tilde{\cal M}_\tau(q_1,q_2)(-e^2)g_{\tau}^{V_1f_1}g_{\tau}^{V_2f_2}
+{\cal O}(\epsilon) \;.
\ea 
Here $\tau$ denotes the quark chirality and the electroweak couplings
$g_{\tau}^{Vf}$ follow the notation of Ref.~\cite{HZ}, with, e.g.,
$g_\pm^{\gamma f}=Q_f$, the fermion electric charge in units of $|e|$,
$g_-^{Wf}=1/(\sqrt{2}\sin\theta_W)$ and $g_-^{Zf}=(T_{3f}-Q_f\sin^2\theta_W)/
(\sin\theta_W\cos\theta_W)$,
where $\theta_W$ is the weak mixing angle and $T_{3f}$ is the third component
of the isospin of the (left-handed) fermions.

A finite contribution of the virtual diagrams, which is proportional 
to the Born amplitude (the $c_{\rm virt}$ term), is pulled out in
correspondence with Eq.~(\ref{eq:vertexvirt}). The remaining 
non-universal term, $\tilde{\cal M}_\tau(q_1,q_2)$, is also finite and 
can be expressed in terms of the finite parts of the Passarino-Veltman 
$B_0$, $C_0$ and $D_{ij}$ functions, which we denote as $\tilde B_0$, 
$\tilde C_0$ and $\tilde D_{ij}$. 
Analytical expressions for these functions, along with the expression 
for $\tilde{\cal M}_\tau(q_1,q_2)$, are given in Appendix~\ref{sec:appendix}.

An equivalent form for Eq.~(\ref{eq:boxlinefig}) has been derived
where all the $\tilde D_{ij}$ have been reduced to $\tilde B_0$, $\tilde C_0$
and $\tilde D_{0}$ functions. We have checked numerically that the two
expressions agree within the numerical precision of the two FORTRAN codes.

The factorization of the divergent contributions to the virtual subamplitudes,
as multiples of $\MB^{(i)}$,
implies that the overall infrared and collinear divergence multiplies
the complete Born amplitude (the sum of the Feynman
graphs of Fig.~\ref{fig:feynBorn}). We can summarize this result for the
virtual corrections to the upper fermion line by writing the complete virtual
amplitude ${\cal M}_V$  as 
\ba
\label{eq:box_tri_contrib}
{\cal M}_V &=& {\cal M}_B\frac{\alpha_s(\mu_R)}{4\pi} C_F
\(\frac{4\pi\mu_R^2}{Q^2}\)^\epsilon \Gamma(1+\epsilon)
\lq-\frac{2}{\epsilon^2}-\frac{3}{\epsilon}+c_{\rm virt}\rq \nonumber\\
&+&\frac{\alpha_s(\mu_R)}{4\pi} C_F(-e^2)\Bigl[ 
\tilde{\cal M}_\tau(q_1,q_2)g_{\tau}^{V_1f_1}g_{\tau}^{V_2f_2} +
\tilde{\cal M}_\tau(q_2,q_1)g_{\tau}^{V_2f_1}g_{\tau}^{V_1f_2} \Bigr]
+{\cal O}(\epsilon) \nonumber \\
&=& {\cal M}_B\frac{\alpha_s(\mu_R)}{4\pi} C_F
\(\frac{4\pi\mu_R^2}{Q^2}\)^\epsilon \Gamma(1+\epsilon)
\lq-\frac{2}{\epsilon^2}-\frac{3}{\epsilon}+c_{\rm virt}\rq 
+\tilde{\cal M}_V \;,
\ea 
where $\tilde{\cal M}_V$ is finite. The interference contribution in the 
cross-section calculation is then given by 
\beq    
\label{eq:virtual_born}
2 \Re \lq {\cal M}_V\MB^* \rq
= |\MB|^2 \frac{\alpha_s(\mu_R)}{2\pi} C_F
\(\frac{4\pi\mu_R^2}{Q^2}\)^\epsilon \Gamma(1+\epsilon)
\lq-\frac{2}{\epsilon^2}-\frac{3}{\epsilon}+c_{\rm virt}\rq\
+2 \Re \lq \tilde{\cal M}_V\MB^* \rq \;.
\eeq
This expression replaces the analogous result for the
NLO corrections to $qq\to qqH$, Eq.~(2.11) in Ref.~\cite{Figy:2003nv}.
The divergent piece appears as the same multiple of the Born amplitude
squared as in the $qq\to qqH$ cross section. It cancels explicitly 
against the phase-space integral of the dipole terms (see Ref.~\cite{CS} and
Eq.~(2.10) of Ref.~\cite{Figy:2003nv})
\beq
\label{eq:I}
\langle \I(\ep) \rangle = |\MB|^2 \frac{\alpha_s(\mu_R)}{2\pi} C_F
\(\frac{4\pi\mu_R^2}{Q^2}\)^\epsilon \Gamma(1+\epsilon)
\lq\frac{2}{\epsilon^2}+\frac{3}{\epsilon}+9-\frac{4}{3}\pi^2\rq\;,
\eeq
which absorbs the real-emission singularities. After this cancellation, all
remaining integrals are finite and can, hence, be evaluated in $d=4$ 
dimensions. This means that the values  of $\MB$ and $\tilde{\cal M}_V$
need to be computed in 4 dimensions only and we use the amplitude 
techniques of Ref.~\cite{HZ} to obtain them numerically.

\subsection{Checks}

We have verified, both analytically and numerically, the gauge invariance of
Eq.~(\ref{eq:box_tri_contrib}): once the extra \mbox{$\eps_1\cdot q_1$} and
$\eps_2\cdot q_2$ terms have been re-inserted in this expression, the 
individual finite subamplitudes $\tilde{\cal M}_\tau(q_i,q_j)$ 
vanish upon the replacements $\eps_1\,\to\,q_1$ or $\eps_2\,\to\, q_2$.
This is a strong check of the tensor reduction and manipulation of the
virtual contributions depicted in Fig.~\ref{fig:boxline}.

We have taken the Born amplitudes for $W$ and $Z$ production from
Ref.~\cite{Chehime:1992ub} and use the real-emission amplitudes of
Ref.~\cite{Rainwater:1996ud} for $Z$ production. In addition,
the $Zjj$ results at the Born level were successfully checked with COMPHEP
code~\cite{ilyin}. 
For $W$ production, the real-emission amplitudes were obtained by modifying 
the previously tested $Zjjj$ amplitudes~\cite{Rainwater:1996ud}. 
We have generated equivalent amplitudes 
with MadGraph~\cite{madgraph}, checking their consistency numerically.

For the $W^{+}$ case, we have built two totally-independent codes.  This has
allowed us to check the overall structure of the dipole-formalism terms
(common to all the vector-boson fusion processes), and to compare tree-level,
real-emission and virtual amplitudes.  The two codes agree within the
numerical precision of the two FORTRAN programs for the total cross sections
and for final-state kinematic distributions.

\section{The parton-level Monte Carlo}
\label{sec:MC}

The cross-section contributions discussed above have been implemented in a 
parton-level Monte Carlo program for $\ell^+\nul jj$, $\ell^-\bar\nul jj$ and
$\ell^+\ell^-jj$ production at NLO in QCD, which is very similar to the 
program for $Hjj$ production by weak-boson fusion described in 
Ref.~\cite{Figy:2003nv}. As in our previous work, the tree-level and the 
finite parts of the virtual amplitudes are calculated numerically, using the 
helicity-amplitude formalism of Ref.~\cite{HZ}. The Monte Carlo integration 
is performed with a modified version of VEGAS~\cite{vegas}. While many 
aspects of our present calculation are completely analogous to those 
described in Ref.~\cite{Figy:2003nv}, several
new problems appear for the vector-boson production processes which require
explanation.

In order to deal with $W/Z$ boson decay
\beq
W/Z(p_{\ell_1}+p_{\ell_2}) \, \to \, \ell_1(p_{\ell_1}) + \ell_2(p_{\ell_2})\,,
\eeq
we have to introduce a finite $W/Z$ width, $\Gamma_V$, in the resonant poles
of the $s$-channel weak-boson propagators. However, in the presence of 
non-resonant graphs, like those of  Figs.~\ref{fig:feynBorn}(e) and~(f), this 
introduces changes in a subclass of Feynman graphs only, which 
leads to a violation of electroweak gauge invariance, which is guaranteed for
the zero-width amplitudes. Such non-gauge-invariant finite-width effects
can lead to huge unphysical enhancements at very small photon virtuality
and should be avoided~\cite{Argyres:1995ym}. For the case at hand, 
transverse-momentum cuts on the
two final-state tagging jets (see Sec.~\ref{sec:pheno}) largely eliminate the
dangerous phase-space  
regions with low-virtuality gauge bosons. Nevertheless, a careful handling
of the finite-width effects is called for.

We have accomplished this using two different schemes.\\
{\bf I.}  In {\em the overall-factor scheme}~\cite{Baur:1991pp}, one
 multiplies all the diagrams shown in Fig.~\ref{fig:feynBorn}, and all
 virtual and real-emission contributions as well, by an overall factor
  \beq
  \label{eq:overall_scheme}
      \frac{(p_{\ell_1} + p_{\ell_2})^2 - m_V^2}
	   {(p_{\ell_1} + p_{\ell_2})^2 - m_V^2 + i m_V \Gamma_V}\; ,
  \eeq
  where $\Gamma_V$ has been assumed to be constant.  This way, close to
  resonance 
  [($p_{\ell_1} + p_{\ell_2})^2 \sim m_V^2$], where the sum of the diagrams is
  dominated by the vector-boson propagator,
  %$1/[(p_{\ell_1} + p_{\ell_2})^2 -  m_V^2]$, 
  we recover the result of the resonance approximation.
  Away from resonance, and, thus, in a subdominant phase-space region,
  the error that we make, by multiplying all the diagrams
  by the factor in Eq.~(\ref{eq:overall_scheme}), is of the order of
  $\Gamma_V/m_V\approx 2.7\%$, for both $Z$ and $W$ boson production.
  
  The advantage of this scheme is that it preserves full $SU(2)\times U(1)$
  gauge invariance, since
  the gauge-invariant set of zero-width diagrams is multiplied by an overall
  factor.

{\bf II.} In {\em the complex-mass scheme}~\cite{Denner:1999gp}, one 
  globally replaces $m_V^2 \rightarrow m_V^2 -i m_V\Gamma_V$, also in the 
  definition of the weak mixing angle, $\sin^2\theta_W=1-m_W^2/m_Z^2$.
  We have implemented a modified complex-mass scheme where we replace 
  $m_V^2 \rightarrow m_V^2 -i m_V\Gamma_V$ in the weak-boson propagators 
  appearing in Fig.~\ref{fig:feynBorn}, but we keep a real value for
  $\sin^2\theta_W$. 
  With this prescription, the electromagnetic Ward identity relating the 
  tree-level triple-gauge-boson vertex, 
  $-ie\Gamma_{WW\gamma}^{\alpha\beta\mu}$, and the inverse $W$ propagator,
  $(D_W)^{-1}_{\alpha\beta}(q)$, is preserved~\cite{lopez}
\begin{equation}
(q_1-q_2)_\mu \Gamma_{WW\gamma}^{\alpha\beta\mu} = 
i(D_W)^{-1}_{\alpha\beta}(q_1) -
i(D_W)^{-1}_{\alpha\beta}(q_2) \;.
\end{equation}

This relation removes potential problems with small $q^2$ photon propagators,
where gauge-invariance-violating terms, proportional to $\Gamma_W/m_W$, may
be enhanced by factors $E_T^2/q^2$, where the hard scale $E_T$ is set by
typical transverse momenta of the process. The corresponding enhancement for
$Z$-boson propagators is of order $E_T^2/(|q^2|+m_Z^2)$ and, hence, small for
the energies available at the LHC. Also, we note that the imaginary part of
$\sin^2\theta_W=1-(m_W^2-im_W\Gamma_W)/(m_Z^2-im_Z\Gamma_Z)$, in the full
complex-mass scheme, is 200 times smaller than the real part and hence well
below the naive expectation $\Gamma_V/m_V\approx 2.7\%$ for the size of
finite-width corrections.

We have used the two different schemes to compute total cross sections 
with VBF cuts and find agreement at the level of the 0.5\% or better.
This ambiguity, thus, represents a minor contribution to 
higher-order electroweak corrections. 

Inspection of the Feynman graphs of Fig.~\ref{fig:feynBorn} shows that
the non-abelian triple-gauge-boson vertices (TGV) enter via the $WWZ$ and 
$WW\gamma$ couplings in diagrams like Fig.~\ref{fig:feynBorn} (d). These 
graphs receive QCD vertex corrections only 
and, therefore, factorize according to Eq.~(\ref{eq:vertexvirt}) in terms 
of the tree-level TGV graphs, independent of the form of the TGV. In
particular, 
the presence of anomalous $WWZ$ or $WW\gamma$ couplings can easily be 
taken into account by a simple modification of the Born amplitude. Our
program supports anomalous couplings $\kappa_\gamma$, $\kappa_Z$, 
$\lambda_\gamma$, $\lambda_Z$
etc.~\cite{Hagiwara:1986vm} and thus allows to extend the analysis of
anomalous-coupling effects in vector-boson fusion processes~\cite{Baur:1993fv}
to NLO QCD accuracy.

The requirement of two observable jets, of finite transverse momentum (see
Sec.~\ref{sec:pheno}), is 
sufficient to render the LO cross section for EW $Wjj$ and $Zjj$ events 
finite. At NLO, initial-state collinear singularities appear. For
$g\to q\bar q$ and $q\to qg$ splitting, these are properly taken into 
account via the renormalization of quark and gluon distribution functions. An 
additional collinear divergence  exists, however, because of the presence
of $t$-channel photons in tree-level graphs, such as in
Fig.~\ref{fig:feynBorn} (a, b, d, e). Real-emission corrections lead to Feynman
graphs such as the one shown in Fig.~\ref{fig:feyndrop} (d): the final-state
$d$ and $\bar u$ quarks may lead to observable jets, allowing vanishing 
momentum transfer for the virtual photon and a corresponding collinear
singularity, representing, in the case shown, a QED correction to the 
LO process $g\gamma\to d\bar u W^+$. This singularity would have to be 
absorbed into the renormalization of the photon distribution function 
inside the proton. Alternatively, one may impose a cut, 
$|t|>Q^2_{\gamma,\rm min}$, on the virtuality of the photon and replace 
the missing piece by the $p\gamma\to VjjX$ cross section, folded with the 
appropriate photon density in the proton~\cite{Baur:1991pp,kniehl}. 
We have chosen this latter
approach: all divergent amplitudes are set to zero below
$Q^2_{\gamma,\rm min}=4$~GeV$^2$ and $p\gamma\to VjjX$ is considered to be a 
separate electroweak contribution to $Vjj$ events, which we do not calculate
here.
\\
When imposing typical VBF cuts, with their large-rapidity separation and
concomitant invariant mass of the two tagging jets, the $p\gamma\to VjjX$
contribution to the EW $Vjj$ cross section is quite small. 
For the VBF cuts defined in the next section, with $p_{Tj}>20$~GeV and
a rapidity separation of the two tagging 
jets of $\Delta y_{jj}>4$, the NLO $W^+jj$ 
cross section, for example, increases by a mere 0.2\% when lowering the 
photon cutoff to $Q^2_{\gamma,\rm min}=0.1$~GeV$^2$ from our 4~GeV$^2$ 
default value\footnote{
The finite proton mass provides an absolute lower bound on the photon 
virtuality, $Q^2_\gamma \gsim m_p^2(m_{Vjj}^2/xs)^2$, where $m_{Vjj}$
is the invariant mass of the produced system and $x$ denotes the Feynman 
$x$ of the colored parton in the subprocesses for $p\gamma\to VjjX$. 
We have chosen the lower cutoff of 
$Q^2_{\gamma,\rm min}=0.1$~GeV$^2$ for a very rough simulation
of the resulting finite photon flux.
}. This number increases to
0.7\% for $\Delta y_{jj}>2$. Because these contributions are negligible, we 
have not yet implemented the calculation of this small missing piece
in our program.

In the computation of cross sections and distributions presented below, we have
used the CTEQ6M parton distribution functions (PDFs)~\cite{cteq6} with
$\alpha_s(m_Z)=0.118$ 
for all NLO results and CTEQ6L1 parton distributions 
%with $\alpha_s(m_Z)=0.130$ 
for all LO cross sections. The CTEQ6 fits include $b$ quarks as an active 
flavor. For consistency, the $b$ quark is included
as an initial- and/or final-state massless parton in all neutral-current 
processes, i.e., we include only those processes with external $b$ quarks,
where no internal top-quark propagator appears via the  $btW$ vertex, being
forbidden by Feynman rules.
Top-quark contributions, obviously, go beyond our massless-fermion 
approximation and would have to be treated as a separate
process. Allowed
neutral-current processes with $b$ quarks appear for $Z$ production only.  
The $b$-quark contributions are quite small, however, affecting the 
$Z$-boson production cross section at the $1$\% level only.

We choose $m_Z=91.188$~GeV, $m_W=80.419$~GeV and the measured value of $G_F$
as our electroweak input parameters, from which we obtain
$\alpha_{QED}=1/132.51$ and $\sin^2\theta_W=0.2223$, using LO electroweak
relations. The decay widths are then calculated as $\Gamma_W=2.099$~GeV and
$\Gamma_Z=2.510$~GeV, which agrees with their Particle Data Group~\cite{PDG}
values at the level of 0.9\% and 0.6\% respectively, which is better
than the overall theoretical uncertainty we are striving for.

In order to reconstruct jets from the final-state partons, the
$k_T$ algorithm~\cite{kToriginal}, as described in Ref.~\cite{kTrunII}, is
used, with resolution parameter $D=0.8$.

\section{Results for the LHC}
\label{sec:pheno}

The parton-level Monte Carlo program described in the previous section
has been used to determine the size of the NLO QCD corrections to EW $Vjj$ 
cross sections at the LHC.
Using the $k_T$ algorithm, we calculate the partonic cross sections for events
with at least two hard jets,  which are required to have
\beq
\label{eq:cuts1}
p_{Tj} \geq 20~{\rm GeV} \, , \qquad\qquad |y_j| \leq 4.5 \, .
\eeq
Here $y_j$ denotes the rapidity of the (massive) jet momentum which is 
reconstructed as the four-vector sum of massless partons of 
pseudorapidity $|\eta|<5$. The two reconstructed jets of highest transverse 
momentum are called ``tagging jets'' and are identified with the final-state
quarks which are characteristic for vector-boson fusion processes. 

We consider decays $Z\to \ell^+\ell^-$ and $W\to \ell\nul$
into a single generation of leptons. In order to ensure that the charged 
leptons are well observable, we impose the lepton cuts
\beq
\label{eq:cuts2}
p_{T\ell} \geq 20~{\rm GeV} \,,\qquad |\eta_{\ell}| \leq 2.5  \,,\qquad 
\triangle R_{j\ell} \geq 0.4 \, ,
\eeq
where $R_{j\ell}$ denotes the jet-lepton separation in the rapidity-azimuthal
angle plane. In addition, the charged leptons are required to fall between
the rapidities of the two tagging jets,
\beq
\label{eq:cuts3}
y_{j,min}  < \eta_\ell < y_{j,max} \, .
\eeq

We do not specifically require the two tagging jets to reside in opposite 
detector hemispheres for the present analysis. 
Backgrounds to VBF are significantly suppressed by requiring
a large rapidity separation of the two tagging jets.
%, which will be called 
%the ``rapidity gap cut'' in the following. 
Unless stated otherwise, we
require
\beq
\label{eq:cuts4}
\Delta y_{jj}=|y_{j_1}-y_{j_2}|>4\; .
\eeq

\begin{figure}[htb] 
%\vspace*{-0.3in}
\centerline{ 
\epsfig{figure=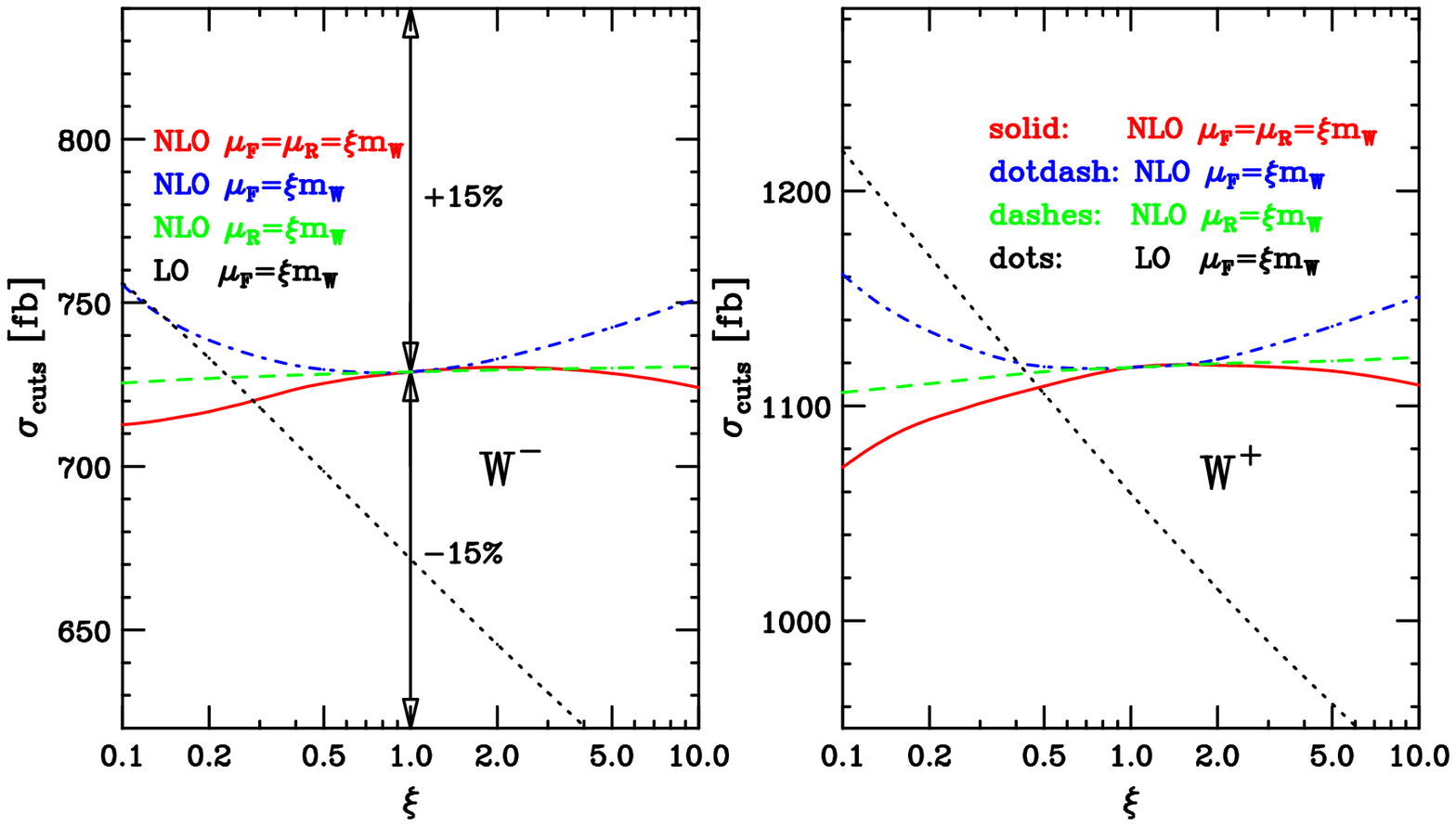,width=0.9\textwidth,clip=}
} 
\ccaption{} 
{\label{fig:scale_depW} 
Scale dependence of the total cross section at LO and NLO within the cuts of
Eqs.~(\ref{eq:cuts1})--(\ref{eq:cuts4}) for $W^-$ and $W^+$ production at 
the LHC. The decay branching ratio of the $W$ is included in the definition
of the cross section, here and in all subsequent figures.
The factorization scale $\mu_F$ and/or the renormalization scale $\mu_R$
have been taken as multiples of the vector-boson mass, $\xi\, m_W$, and
$\xi$ is varied in the range $0.1 < \xi < 10$. The NLO curves are for
$\mu_F=\mu_R=\xi m_W$ (solid red line), 
$\mu_F=m_W$ and $\mu_R=\xi\, m_W$ (dashed green line) and 
$\mu_R=m_W$ and $\mu_F$ variable (dot-dashed blue line).
The dotted black curve shows the dependence of the LO cross
section on the factorization scale. At this order, $\alpha_s(\mu_R)$ does not 
enter.
}
\end{figure}

\begin{figure}[thb] 
%\vspace*{-0.3in}
\centerline{ 
\epsfig{figure=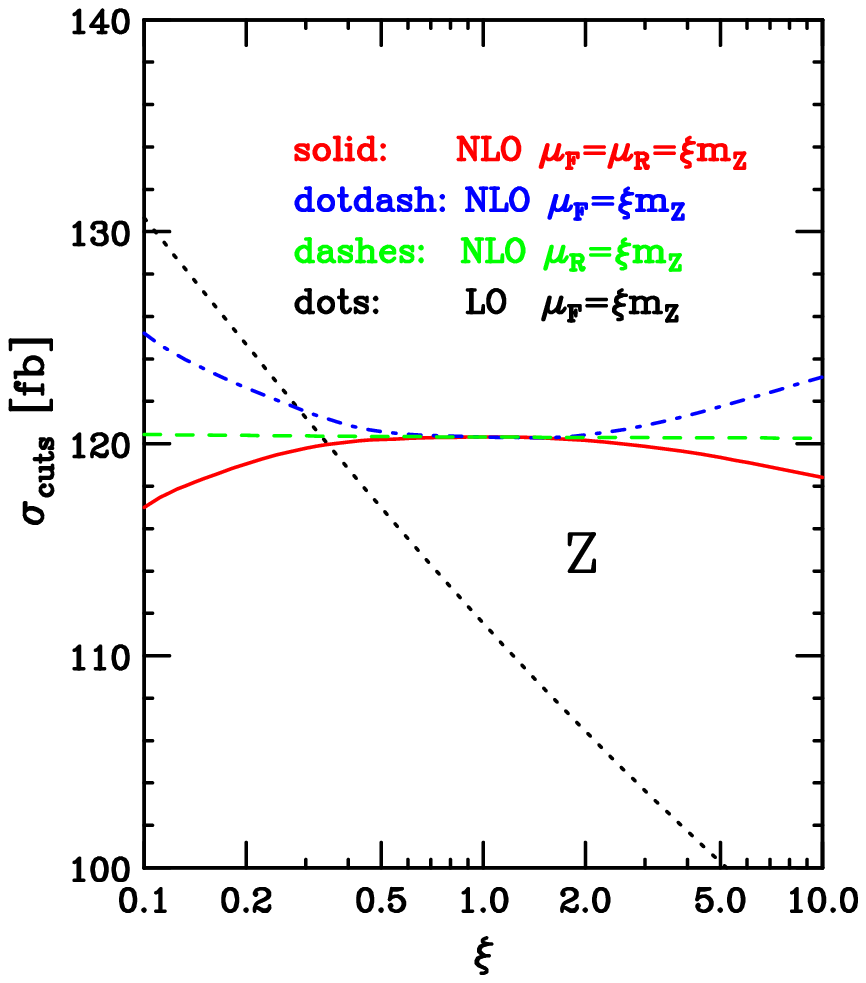,width=0.522\textwidth,clip=}%
%\epsfig{figure=scale_dep_Z.ps,angle=90,width=0.522\textwidth,clip=}
% \ \
%\hspace{0.478\textwidth}
} 
\ccaption{} 
{\label{fig:scale_depZ} 
Same as  Fig.~\ref{fig:scale_depW}, but for $Z$ production at the LHC, with
the $Z\to \mu^+\mu^-$ branching ratio included in the definition of the cross
section, here and in all subsequent figures.
}
\end{figure}

Cross sections, within the cuts of Eqs.~(\ref{eq:cuts1})--(\ref{eq:cuts4}), 
are shown in Fig.~\ref{fig:scale_depW}, for $Wjj$ production, and in
Fig.~\ref{fig:scale_depZ}, for the $Zjj$ case. In both figures, the scale 
dependence of the LO and NLO cross sections is 
shown for fixed renormalization and factorization scales, $\mu_R$ and
$\mu_F$, which are tied to the masses of the produced vector bosons $m_V$
\bq
\label{eq:scale.mV}
\mu_R = \xi_R\,m_V\;,\qquad\qquad \mu_F = \xi_F\,m_V\; .
\eq
The LO cross sections only depend on $\mu_F=\xi\,m_V$. At NLO we show
three cases: (a) $\xi_F=\xi_R=\xi$ (red solid line); (b) $\xi_F=\xi$,
$\xi_R=1$ (blue dot-dashed line); and (c) $\xi_R=\xi$, $\xi_F=1$ 
(green dashed line). While the factorization-scale dependence of the LO result
is sizable, the NLO cross sections are quite insensitive to scale variations:
allowing a factor 2 variation in either directions, i.e., considering the range
$0.5<\xi <2$, the NLO cross sections change by less than 1\% in all cases.

As a second option, we have considered scales tied to 
the virtuality of the exchanged electroweak bosons. Specifically, independent 
scales $Q_i$ are determined as in 
Eqs.~(\ref{eq:vertexvirt}) and~(\ref{eq:boxlinefig})  
for radiative corrections on the upper and on the 
lower quark line, and we set 
\beq
\label{eq:scale.Q}
\mu_{Fi}=\xi_F Q_i\; ,\qquad\qquad \mu_{Ri}=\xi_R Q_i\;.  \eeq This choice is
motivated by the picture of VBF as two independent deep-inelastic scattering
type events, with independent radiative corrections on the two
electroweak-boson vertices. Resulting $Vjj$ cross sections at NLO are about
1\% lower for $\mu_F=\mu_R=Q_i$ than for $\mu_F=\mu_R=m_V$. In the following,
we refer to the latter choice as the ``$M$ scheme'' while the choice
$\mu_F=\mu_R=Q_i$ is called the ``$Q$ scheme''. As we will see below, a
residual NLO scale dependence of about 1\%--2\% is also typical for
distributions, resulting in very stable NLO predictions for $Vjj$ cross
sections.

In addition to these quite small scale uncertainties, we have estimated the
error of the $W^\pm jj$ cross sections due to uncertainties in the
determination of the PDFs. This error is
determined by calculating the total $Wjj$ cross section, within the cuts of
Eqs.~(\ref{eq:cuts1})--(\ref{eq:cuts4}), using two different sets of PDFs
with errors, computed by the CTEQ~\cite{cteq6} and MRST~\cite{mrst}
Collaborations.
Together with the PDF that gives the best fit to the data, the CTEQ6M set
provides 40 PDFs, and the MRST2001E  30 PDFs, which correspond
to extremal plus-minus variations in the directions of the error eigenvectors
of the Hessian, in the space of the fitting parameters.
To be on the conservative side, we have added the maximum deviations for each
error eigenvector in quadrature, and we have found a total PDF uncertainty of
$\pm 4\%$ with the CTEQ PDFs, and of roughly $\pm 2\%$ with the MRST set.

\begin{figure}[thb] 
%\vspace*{-0.3in}
\centerline{ 
\epsfig{figure=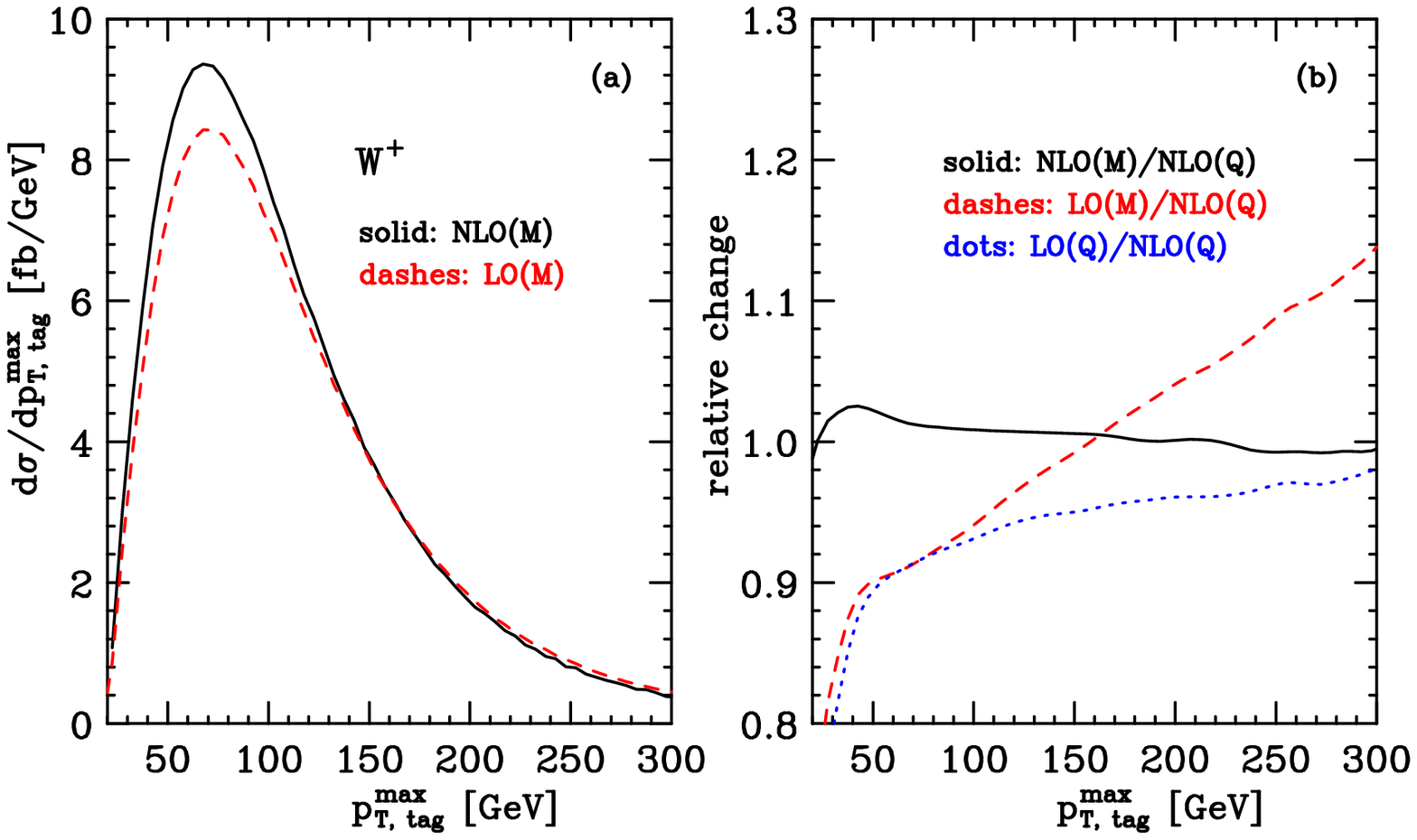,width=0.9\textwidth,clip=} 
} 
\ccaption{} 
{\label{fig:pt_max_tagj}
Transverse-momentum distribution of the highest-$p_T$ 
tagging jet in $W^+$ production at the LHC. In panel~(a) the NLO result
(solid black line) and the LO curve (dashed red line) are shown for the scale
choice $\mu_F=\mu_R=m_W$ ($M$ scheme).
In panel~(b), we show the ratios of the NLO differential cross section in
the $M$ scheme (solid black line), of the LO one in the $M$ scheme (dashed
red line) and of the LO one in the $Q$ scheme (blue dotted line) 
to the NLO distribution in
the $Q$ scheme, which is defined via the scale choice $\mu_F=\mu_R=Q_i$.
}
\end{figure}

\begin{figure}[htb] 
%\vspace*{-0.3in}
\centerline{ 
\epsfig{figure=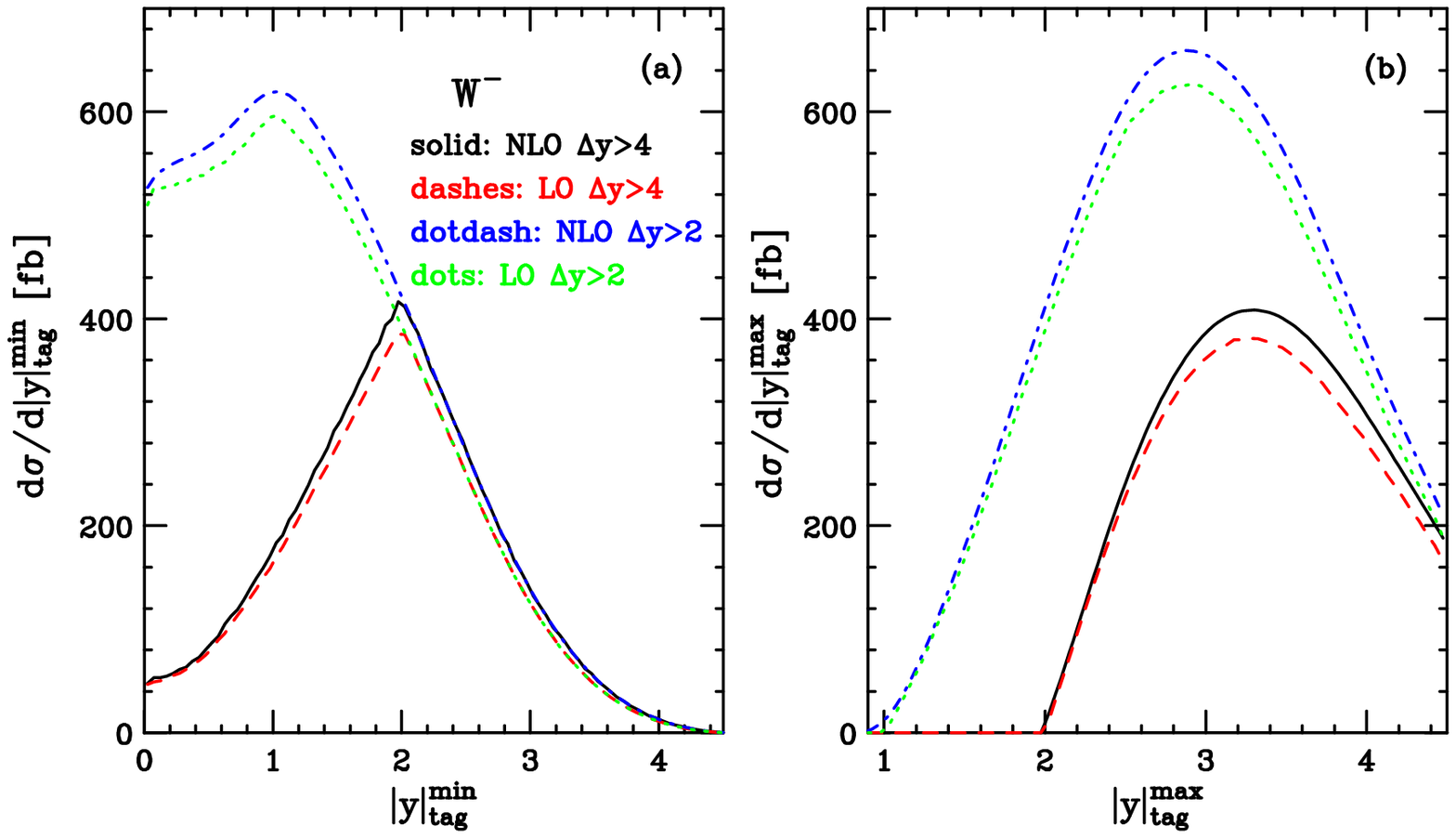,width=0.9\textwidth,clip=} 
} 
\ccaption{} 
{\label{fig:y_tags} 
$W^-$ production cross section as a function of~(a) the smaller and~(b) the
larger absolute value of the two tagging-jet rapidities.  Results are shown
for a rapidity separation between the two tagging jets greater than 2 and 4
(higher and lower pairs of curves, respectively). The LO cross section is
always slightly below the NLO result. Due to the rapidity cut of
Eq.~(\ref{eq:cuts1}), the distributions are truncated at $|y_j| =4.5$.
}
\end{figure} 

For precise comparisons with future LHC data, the residual theoretical error
on the jet and lepton distributions must be estimated. As a first example,
we show the transverse-momentum distribution of the highest-$p_T$ tagging jet
for $W^+jj$ production 
in Fig.~\ref{fig:pt_max_tagj} (a): the shape of the $p_T$ distribution is
fairly similar at LO (red dashed curve) and NLO (black solid line). Both
curves were obtained with a scale choice of $\mu_R=\mu_F=m_W$. In the 
right-hand panel their ratio to the NLO curve with $\mu_R=\mu_F=Q_i$ is 
shown. The ratio of the two NLO distributions deviates from unity by 2\% or 
less over the entire range, which, again, points to the small QCD dependence
of our calculation. 
\begin{figure}[!thb] 
%\vspace*{-0.3in}
\centerline{ 
\epsfig{figure=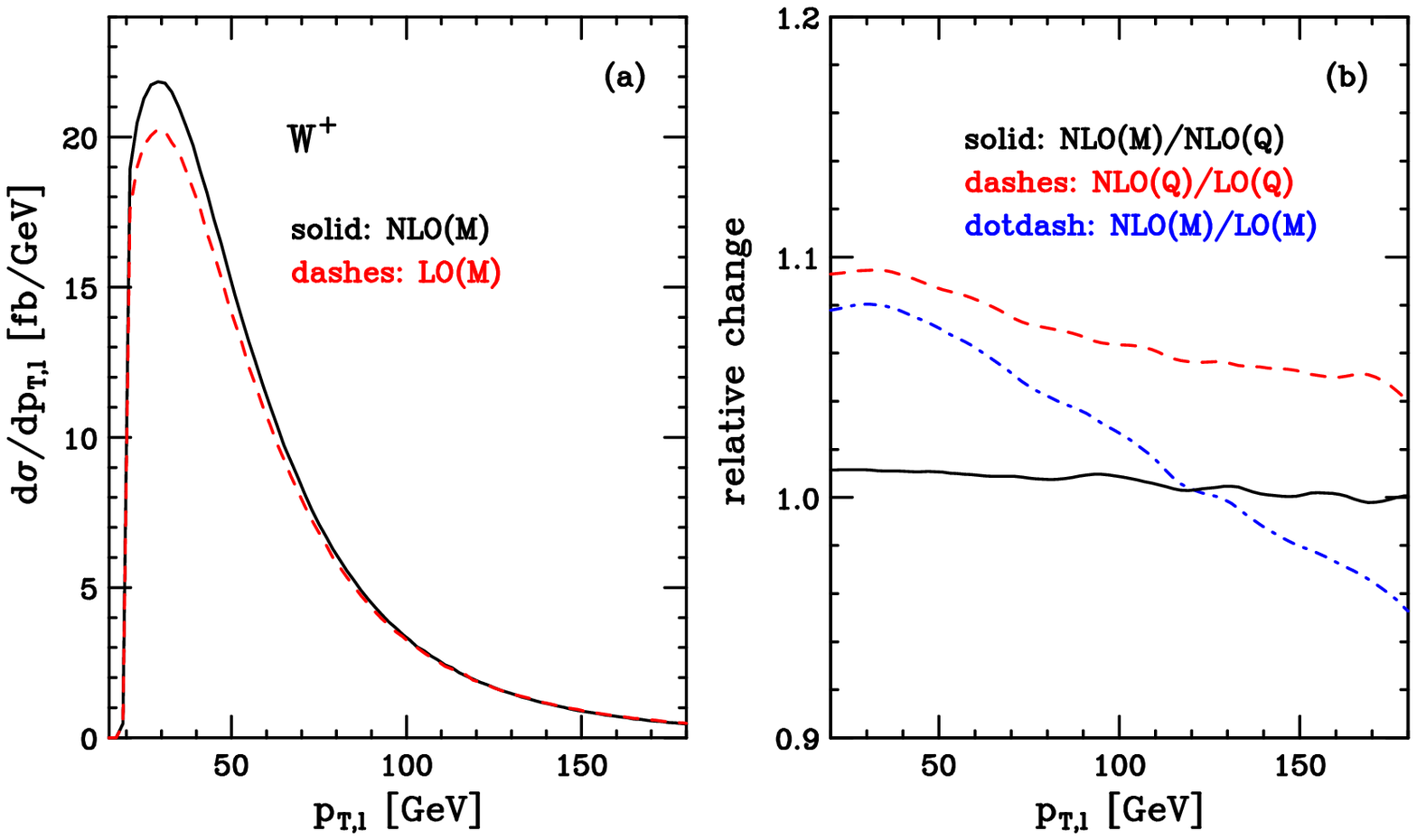,width=0.88\textwidth,clip=} 
} 
\centerline{ 
\epsfig{figure=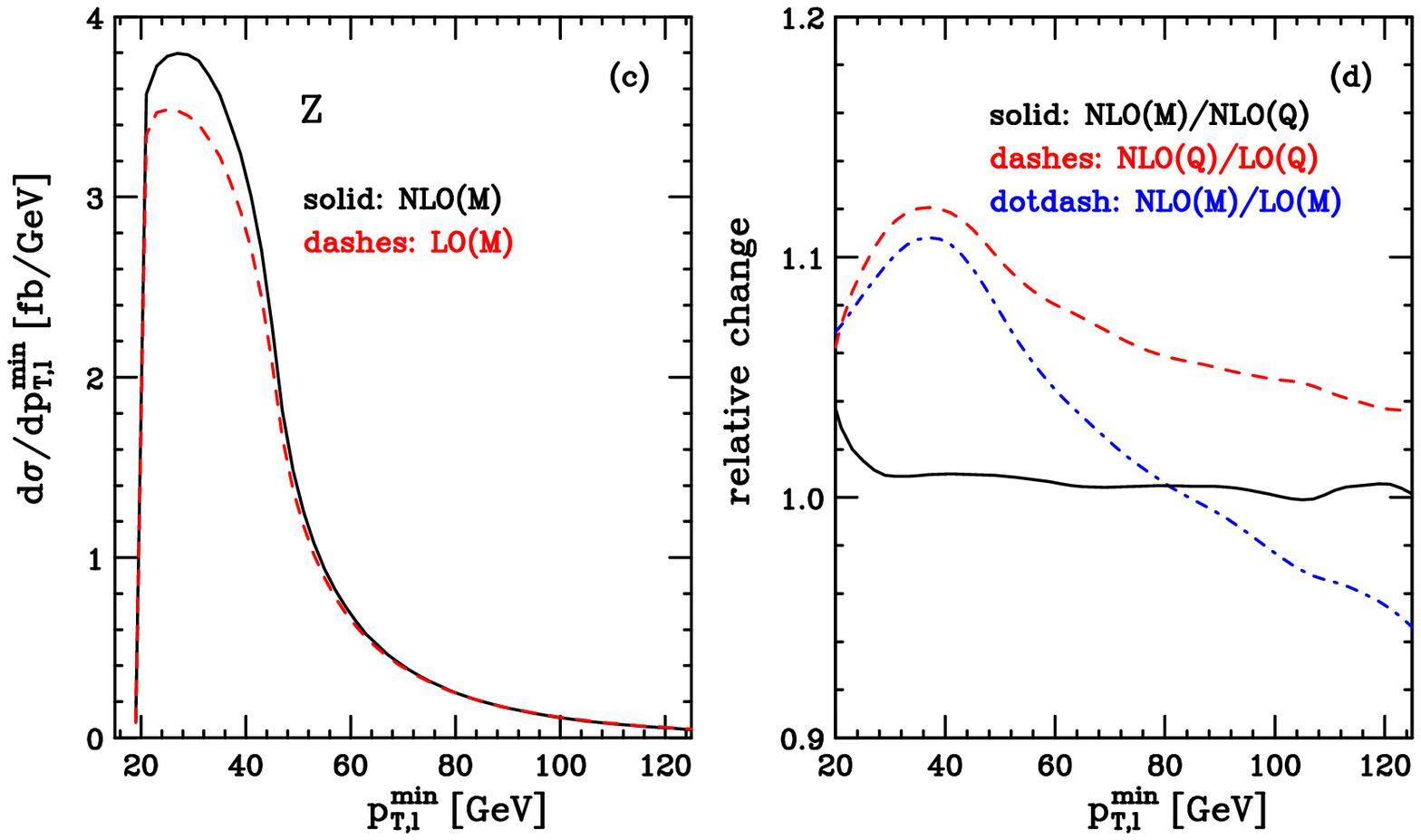,width=0.88\textwidth,clip=} 
} 
\ccaption{} 
{\label{fig:pt_L_min}
Transverse-momentum distributions of the charged final-state lepton in $W^+$
production [panels~(a) and ~(b)] and of the softest of the two final-state
leptons in $Z$ production [panels~(c) and~(d)].  The solid black curves in
panel~(a) and~(c) represent the NLO cross sections and the red dashed curves
the LO ones, for scales $\mu_R=\mu_F=m_V$ ($M$ scheme). Panels~(b) and~(d)
show the ratio of the NLO transverse-momentum distribution 
computed in the $M$ and $Q$ scheme (black solid line), and the $K$ factors in
the $Q$ (red dashed line) and $M$ (blue dot-dashed line) schemes.
}
\end{figure} 

In contrast to the stability of the NLO result, the LO curves depend 
appreciably on the scale choice. The blue dotted line and the red dashed line
in Fig.~\ref{fig:pt_max_tagj} (b) give the ratio of the LO curves for 
$\mu_F=Q_i$ and $\mu_F=m_W$, respectively,
to the NLO result. The shape of the LO
curves, in particular for a constant scale choice like $\mu_F=m_W$, is quite
different from the more reliable NLO result. For transverse-momentum 
distributions we generally find that the ``dynamical'' scale choice 
$\mu_F=Q_i$, at LO, 
better reproduces the shape of the NLO distributions, and is thus 
preferable to a fixed scale. At NLO, or higher order, where the definition
of the momentum transfer $Q_i$ becomes more problematic, the fixed-scale
choice becomes more natural. However, because of the greater stability of
the cross-section prediction, the scale selection also becomes less of a 
phenomenological issue.

Rapidity distributions of the two tagging jets are shown in 
Fig.~\ref{fig:y_tags}, at LO and NLO, and for two
choices of the rapidity-gap requirement, $\Delta y_{jj}>2$ and  
$\Delta y_{jj}>4$. The shapes of the rapidity distributions for the more 
central tagging jet, panel~(a), and the more forward tagging 
jet, panel~(b), are quite similar at LO and NLO. 
In fact, the $K$ factors for these distributions are fairly flat, and 
adequately described by a constant value of about 1.1.
The results in Fig.~\ref{fig:y_tags} were obtained for a fixed scale 
$\mu_F=\mu_R=m_W$ and are for $W^-jj$ production. Curves for the $W^+jj$
and $Zjj$ cross sections are very similar in shape and show the preservation
of shape between LO and NLO curves.

While tagging-jet distributions are quite similar for electroweak $Wjj$ 
and $Zjj$ events at the LHC, the presence of two charged leptons in the 
$Zjj$ case results in somewhat more noticeable differences. 
When considering changes in the lepton $p_T$ cut of Eq.~(\ref{eq:cuts2}),
the transverse momentum of the softer lepton is critical for $Z$ production,
while the single charged lepton must be considered for $Wjj$ events.
These distributions are shown in 
Fig.~\ref{fig:pt_L_min} for $W^+$ production (top panels) and $Z$ production
(bottom panels). 
At NLO the scale variations are again very small, at the 1\%
level, as demonstrated by the ratios of the NLO $p_T$ distributions for
$\mu_F=\mu_R=m_V$ and $\mu_F=\mu_R=Q_i$ (solid black lines) in 
Fig.~\ref{fig:pt_L_min} (b, d). Varying either scales by a factor of 2 leads
to the same conclusion of 1\%--2\% scale uncertainties for the NLO results.
Comparing the LO predictions (dashed and dot-dashed curves) with the 
very precise NLO results shows theoretical errors of the order of 10\%.
Again, as for the jet $p_T$ distributions discussed earlier,
the choice $\mu_F=Q_i$ is better for simulating the shape of the lepton $p_T$ 
distribution at LO. A fixed scale, $\mu_F=m_V$, predicts too steep a fall-off
at large $p_T$. One should note, however, that for the electroweak $Vjj$
processes considered here, these differences are exceptionally small already
at LO: the differences between the LO curves in Fig.~\ref{fig:pt_L_min}
are of the order of 10\% only.

\begin{figure}[!thb] 
%\vspace*{-0.3in}
\centerline{ 
\epsfig{figure=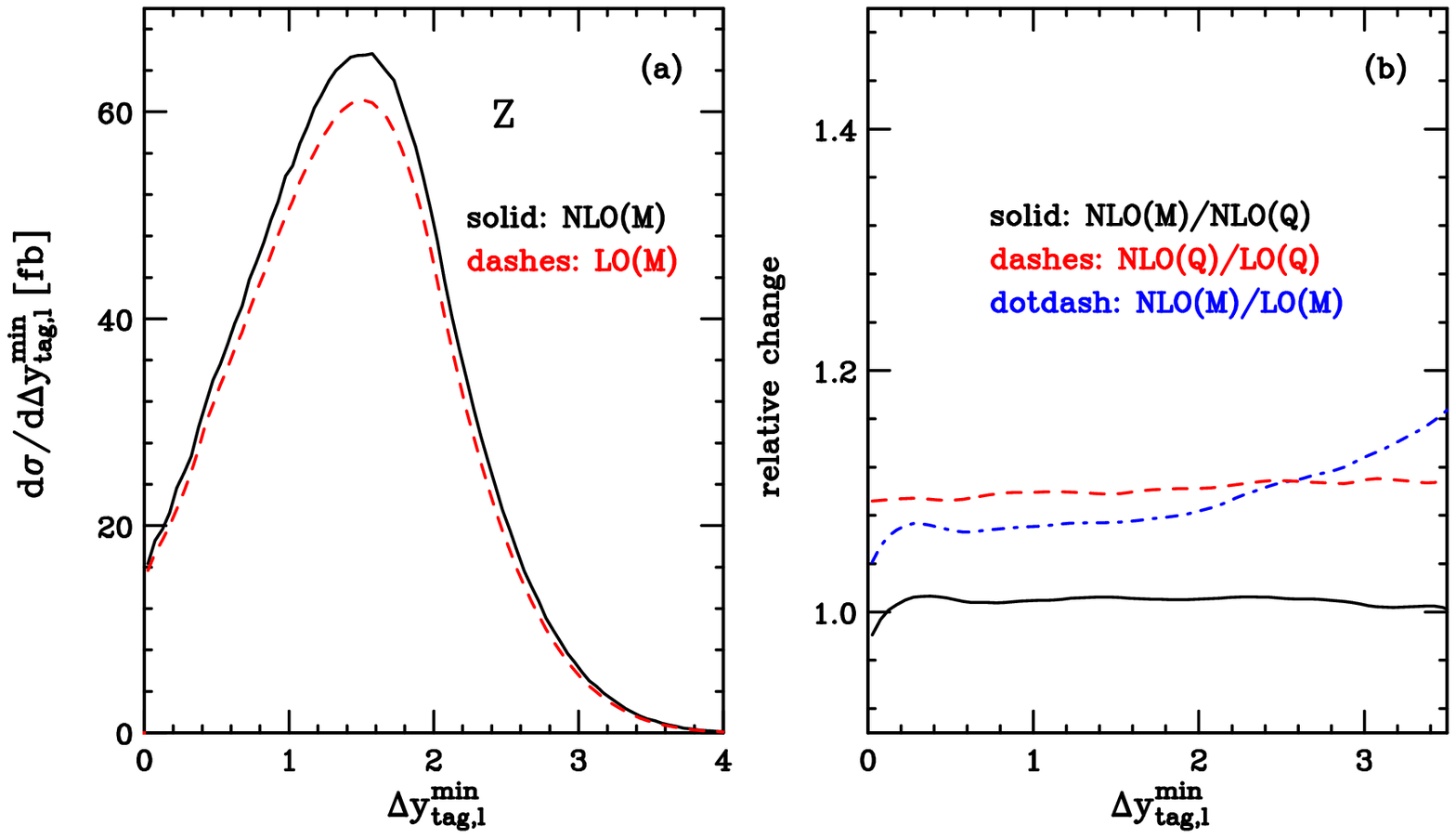,width=0.9\textwidth,clip=}
} 
\centerline{ 
\epsfig{figure=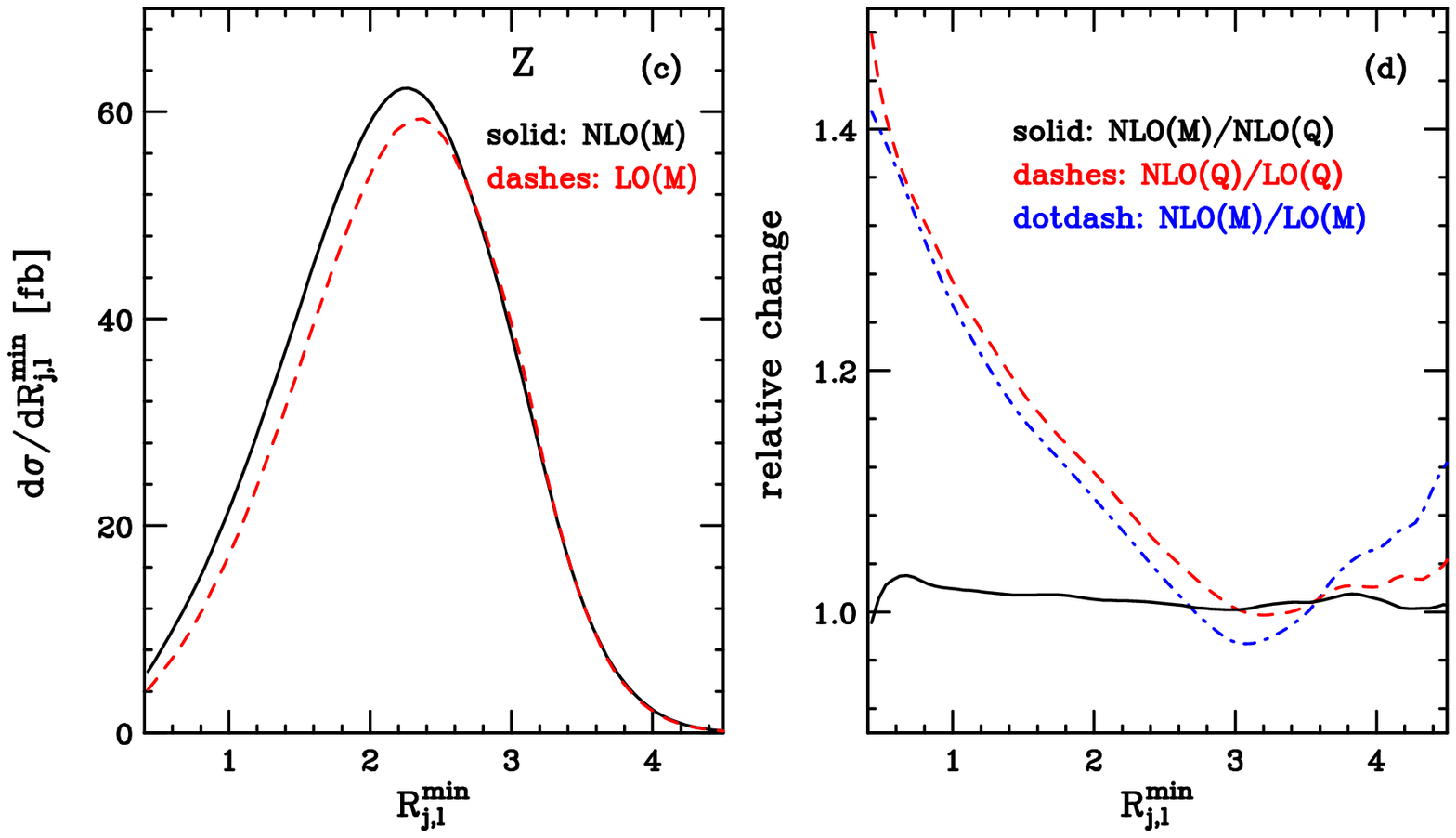,width=0.9\textwidth,clip=}
} 
\ccaption{} 
{\label{fig:delta_y_L_tag_min} 
Angular correlations of leptons and jets in Z production.  Panels~(a) and~(b)
show the minimum rapidity separation between the two leptons and the two
tagging jets. Panels~(c) and~(d) are for the minimum rapidity-azimuthal angle
separations between the leptons and any reconstructed jets (not necessarily
the two tagging jets).  The NLO differential cross sections are shown in
black solid lines, while the LO ones are displayed as red dashed
lines. Scales are fixed in the $M$ scheme.  Panels~(b) and~(d) show the ratio
between the two NLO differential cross sections in the $M$ and $Q$ scheme
(solid black lines) and their respective $K$ factors.  }
\end{figure} 

In contrast to the lepton transverse-momentum distributions described above, 
the shape of the lepton-rapidity distributions is virtually unaffected by 
the NLO corrections: an overall constant $K$ factor is sufficient to describe 
NLO effects. Larger changes are found when considering angular 
correlations of the leptons and jets, which we show for $Zjj$ production 
in Fig.~\ref{fig:delta_y_L_tag_min}. The top panels show the minimal 
rapidity between any of the two leptons and the two tagging jets, 
$\Delta y_{{\rm tag},l}^{\rm min}$. As before,
the tagging jets are taken as the two highest transverse-momentum jets 
in the event ($p_T$ selection). 
The two bottom panels show the minimal separation in the 
rapidity-azimuthal angle plane of the two leptons from any jet (not
necessarily the two tagging jets) in the event,
$R_{j,l}^{\rm min}$.
In both cases, the two scale choices for the NLO result show excellent 
agreement (black solid lines in Fig.~\ref{fig:delta_y_L_tag_min} (b, d)). 
However, the dynamical $K$ factors 
\bq
K(x) = \frac{d\sigma_{NLO}/dx}{d\sigma_{LO}/dx}
\eq
for $x=\Delta y_{{\rm tag},l}^{\rm min}$ and $x=R_{j,l}^{\rm min}$
show qualitatively different behavior. While 
$K(\Delta y_{{\rm tag},l}^{\rm min})$ is fairly constant, i.e., the shape
of the distribution is well described by the LO approximation, the minimal 
lepton-jet separation, $d\sigma/dR_{j,l}^{\rm min}$, shifts noticeably to 
smaller values at NLO. This 
behavior was to be expected, since additional parton emission 
in the higher-order calculation reduces lepton isolation. What is remarkable,
then, is that the selection of the tagging jets as the two highest-$p_T$ jets
does not affect the lepton-tagging jet separation. As for the Higgs boson
case~\cite{Figy:2003nv}, this selection of the tagging jets provides 
excellent correspondence of the LO- and NLO-event topology.

\begin{figure}[thb] 
%\vspace*{-0.3in}
\centerline{ 
\epsfig{figure=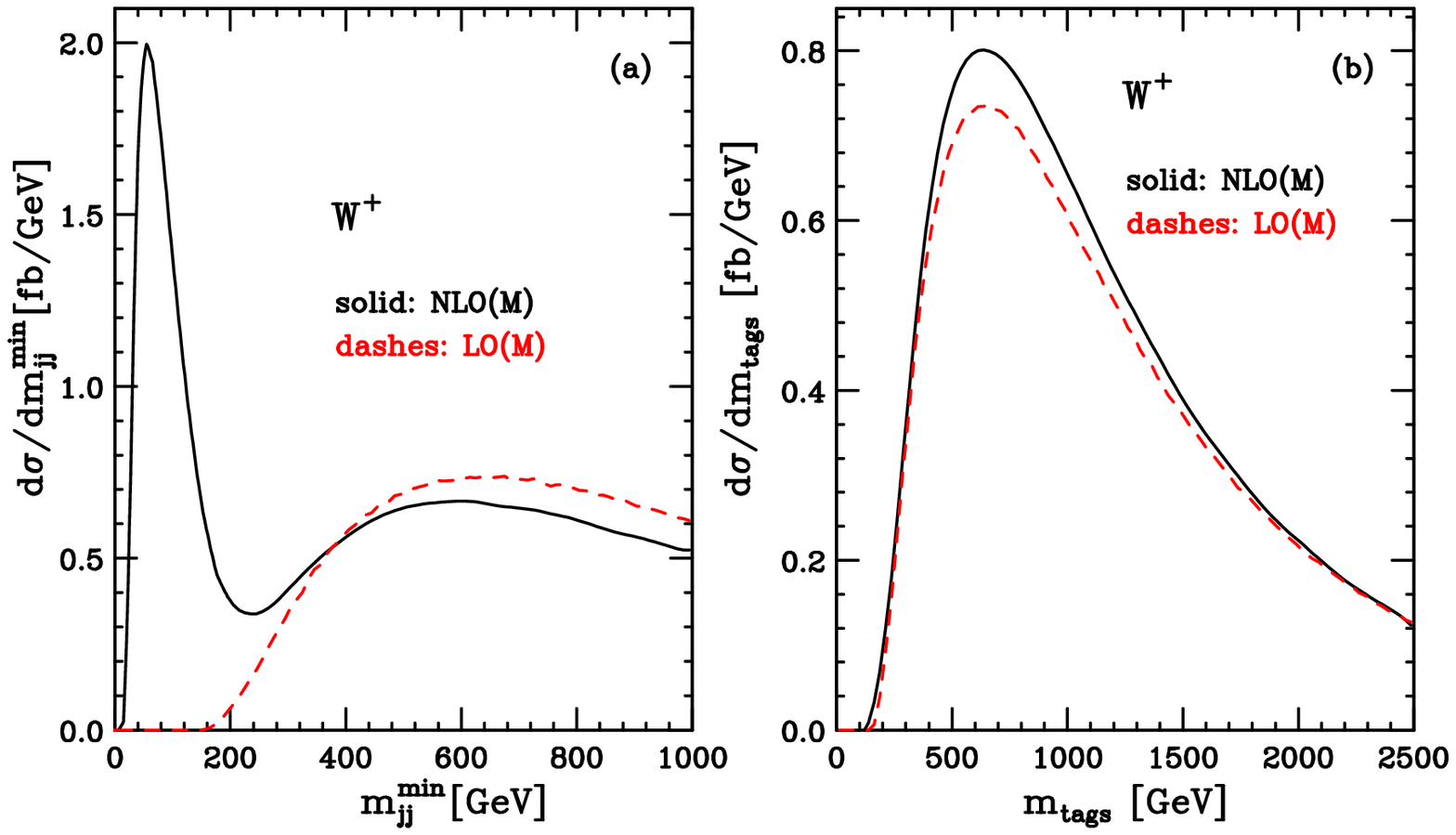,width=0.9\textwidth,clip=} 
} 
\ccaption{} 
{\label{fig:m_jj_min_Wp} 
Dijet invariant-mass distributions for $W^+$ production, with scales in the 
$M$ scheme.
Shown are~(a) the minimum dijet invariant-mass distribution for any 
final-state reconstructed jets (not necessarily the two tagging jets)
and~(b) the invariant mass of the two tagging jets. NLO results are 
shown in solid black lines, while the red dashed lines are for LO
distributions. 
}
\end{figure} 

In order to stress this point we show dijet invariant-mass distributions
for the reconstructed jets (not necessarily the two tagging jets)
for $W^+jj$ events at LO (red dashed lines) and at NLO (solid black lines) in 
Fig.~\ref{fig:m_jj_min_Wp}. The distribution with respect to the minimal 
dijet invariant mass in the event is shown in Fig.~\ref{fig:m_jj_min_Wp} (a)
while Fig.~\ref{fig:m_jj_min_Wp} (b) uses the invariant mass of the two 
tagging jets, $m_{\rm tags}$. 
At LO, there are only two final-state quarks of $p_T>20$~GeV
in each event and, hence, the two curves are identical. At NLO, additional
parton emission provides for soft third jets which form low invariant-mass 
pairs with one of the tagging jets, and this pair shows up as a low-mass
peak in $d\sigma/dm_{jj}^{\rm min}$. Generic selections of the two
tagging jets in a multijet environment tend to pick up some of these 
low-mass pairs and lead to substantial differences in the invariant-mass
distribution of the two tagging jets at LO and at NLO. The $p_T$  
selection of tagging jets, which we have used throughout and for which 
results are shown in Fig.~\ref{fig:m_jj_min_Wp} (b), 
is remarkable in that it preserves the shape of the tagging
jet invariant-mass distribution, $d\sigma/dm_{\rm tags}$, when going from 
LO to NLO.

\section{Conclusions}
\label{sec:summary}
Vector-boson fusion at the LHC represents a class of electroweak processes 
which are under excellent control perturbatively. This has been known for
some time for the most interesting process in this class: 
Higgs boson production via VBF has a modest $K$ factor of about 1.05
for the inclusive production cross section~\cite{Han:1992hr} and this result
also holds when applying realistic acceptance cuts~\cite{Figy:2003nv}. 

In the present paper, we have extended this result to the electroweak
production of $W$ and $Z$ plus two jets, when the final-state particles are
in a kinematic configuration typical of VBF events.
More precisely, we have calculated the NLO QCD corrections to electroweak
production of $\ell \nul jj$ and $\ell^+\ell^- jj$ at LHC, and we have 
implemented them in a fully-flexible NLO Monte Carlo program. $K$ factors are
of the same size as for the Higgs boson production process, typically ranging
between 1.0 and 1.1 for most distributions. What is more important is the
stability of the NLO result: residual scale dependence is at the 2\% level or
below. This is smaller than the present parton-distribution-function
uncertainties, which we have calculated for the $W^\pm jj$ cross sections. We
estimate 4\% PDF errors using CTEQ6M parton distributions and roughly half
that size using MRST2001E PDFs.

Given the excellent theoretical control which we now have for EW $Vjj$
production, these processes can be used as testing grounds for
Higgs boson production in VBF: techniques should be developed to measure
hadronic properties, like forward-jet tagging efficiencies or
central-jet-veto probabilities, in $Wjj$ or $Zjj$ production at the LHC and
to extrapolate these results to Higgs boson production, thus reducing the
systematic errors for Higgs boson coupling measurements. We leave such
applications for the future.

\section*{Acknowledgments}

Part of this work was done at LAPTH in Annecy and D.Z. would like to thank
the members of the laboratoire for their hospitality. 
This research was supported in part by the University
of Wisconsin Research Committee with funds granted by the Wisconsin Alumni
Research Foundation and in part by the U.S.~Department of Energy under
Contract No.~DE-FG02-95ER40896.
C.O. thanks the UK Particle Physics and Astronomy Research Council for
supporting his research.

\appendix
\section{Virtual corrections}
\label{sec:appendix}
In this appendix, we give the expression for the finite, reduced amplitude
$\tilde{\cal M}_\tau(q_1,q_2)$ that appears in Eqs.~(\ref{eq:boxlinefig})
and~(\ref{eq:box_tri_contrib}), in terms of $\tilde B_0$, $\tilde C_0$ and
$\tilde D_{ij}$ functions. Here $\tilde B_0$, $\tilde C_0$ and
$\tilde D_{ij}$ are the finite parts of the Passarino-Veltman $B_0$, $C_0$ 
and $D_{ij}$ functions~\cite{Passarino:1978jh}, and are given explicitly
below. We have also derived $\tilde{\cal M}_\tau(q_1,q_2)$ in terms of
$\tilde B_0$, $\tilde C_0$ and $\tilde D_{0}$ functions, but do not show 
this expression here, due to its length.
We write 
\bq
\label{eq:boxlinedef}
\tilde{\cal M}_\tau(q_1,q_2) = \overline\psi(k_2)\lq 
c_1 \sla\epsilon_1
+c_2\sla\epsilon_2 +c_q\(\sla q_1-\sla q_2\) + 
c_b\sla\epsilon_2\(\sla k_2+\sla q_2\) \sla\epsilon_1 \rq
\frac{1+\tau\gamma_5}{2}\psi(k_1)\,, 
\eq
where $\eps_1=\eps_1(q_1)$ and $\eps_2=\eps_2(q_2)$ are the effective 
polarization vectors of the two electroweak gauge bosons.
The coefficient function $c_1=c_1(q_1,q_2)$ is given by
\ba
c_1 &=&                 %c_1(q_1,q_2) = 
2\epsbkb T_\epsilon\(q_2^2,t\)
-2 \lq \xd{12}+\xd{24}\rq \epsbkb \(q_1^2+q_2^2-3 s-4 t\)
\nonumber\\
    &-&  2 \lq \xd{12}-\xd{24}\rq \epsbqa \(q_2^2-t\) \nonumber\\
    &+& 4 \Bigl[-\xd{11} \epsbkb s - \xd{12} \epsbka t + \xd{13} \epsbkb 
       \(q_2^2-s-t\) \nonumber\\
    &+& \xd{13} \epsbqa q_2^2 - \xd{21} \epsbkb s 
      - \xd{22} \epsbkb t \nonumber\\
    &-& \xd{22} \epsbqa q_2^2 + \xd{23} \epsbkb q_1^2
      + \xd{25} \epsbkb \(q_2^2-s-2 t\) \nonumber\\
    &-& \xd{26} \epsbkb \(q_2^2-s-t\)+\xd{26} \epsbqa t+2 \xd{27} \epsbqa
\nonumber\\
    &-& \xd{32} \epsbkb q_2^2 - \xd{34} \epsbkb (q_2^2-t) 
          \nonumber\\
    &+& \xd{36} \epsbkb \(2 q_2^2-t\) 
      + \xd{37} \epsbkb q_1^2 \nonumber\\
&+& \xd{35} \epsbkb \(q_2^2-s-t\) + \xd{38} \epsbkb
      \(q_1^2+q_2^2-s\)\nonumber\\ 
    &-& \xd{39} \epsbkb q_1^2 
      - \xd{310} \epsbkb \(q_1^2+2 q_2^2-2 s-t\) \nonumber\\
    &-& 4 \xd{311} \epsbkb + 6 \xd{312} \epsbkb + 2 \xd{313} \epsbqa\Bigl]\;,
\ea
where
\bq
T_\epsilon\(q^2,t\) = \frac{1}{t-q^2}\lg
\lq\tilde B_0(t)-\tilde B_0(q^2)\rq \frac{2t+3q^2}{t-q^2}
       +2\tilde B_0(q^2)+1-2q^2\tilde C_0(q^2,t) \rg
\eq
is defined in terms of the finite parts of the $B_0$ and $C_0$ functions
\bq\label{eq:b0tilde}
\tilde B_0(q^2) = 2-\ln\frac{q^2+i0^+}{s}
\eq
and 
\bq\label{eq:c0tilde}
\tilde C_0(q^2,t) = \frac{1}{2(t-q^2)}\biggl(\ln^2\frac{q^2+i0^+}{s}-
\ln^2\frac{t+i0^+}{s}\biggr)\,.
\eq
These expressions are obtained by pulling a common factor 
$\Gamma(1+\eps)(-s)^{-\eps}\equiv{\Gamma(1+\eps)}/(Q^2)^\eps$ 
out of all amplitudes and Passarino-Veltman functions, e.g.,
\beqn
B_0(q^2) &=& \int\frac{d^dk}{i\pi^{d/2}} \frac{1}{k^2(k+q)^2} =
 \frac{\Gamma(1+\eps)}{\eps}
\frac{\Gamma(1-\eps)^2}{\Gamma(2-2\eps)} (-q^2-i0^+)^{-\eps} \nonumber \\
&=& \frac{\Gamma(1+\eps)}{\(-s\)^\eps}
\left[\frac{1}{\eps}+2
-\ln\frac{q^2+i0^+}{s} + \ord{\eps}\right]
= \frac{\Gamma(1+\eps)}{\(Q^2\)^\eps} \left[
\frac{1}{\eps}+\tilde B_0(q^2)+ \ord{\eps} \right] \,. \phantom{aaaaa}  
\eeqn

For the other coefficient functions $c_i=c_i(q_1,q_2)$ we find
\ba
      c_2&=& -2\lq \xd{12}+\xd{24}\rq \lq \epsakb \(q_1^2+q_2^2-s-2 t\)
       +\epsaqb \(q_2^2-s-3 t\)\rq \nonumber\\
 &+&  4 \Big[\xd{13} \epsakb q_1^2 - 
       \xd{13} \epsaka (2 s+t) + \xd{22} \epsaka q_2^2 \nonumber\\
 &-&   \xd{23} \epsakb t + 
         \xd{23} \epsaqb \(q_1^2-t\) - \xd{24} \epsaka q_2^2 \nonumber\\
 &+& \xd{25} \epsakb q_1^2 + \xd{25} \epsaka \(q_2^2-2 s-t\) 
       + \xd{26} \epsakb t \nonumber\\
 &-& \xd{26} \epsaka \(q_1^2-s\) - 
       2 \xd{27} \epsaqb + \xd{33} \epsakb q_1^2\nonumber\\
 &+& \xd{33} \epsaqb q_1^2
       + \xd{37} \epsaka \(q_2^2-s-t\) + \xd{38} \epsaka q_2^2\nonumber\\
 &-& \xd{39} \epsaka \(q_1^2+q_2^2-s\) - \xd{310} \epsaka \(q_2^2-t\)
\nonumber\\
 &+& 2 \xd{311} \epsakb+2 \xd{312} \epsaqb-6 \xd{313} \epsaka\Big]
\nonumber\\
    &+& 2\epsaka T_\epsilon\(q_1^2,t\)\;,
\ea
\ba
   c_q &=& \lq \xd{12}+\xd{24}\rq \epsaepsb s + 
         2 \Big[4 \xd{12} \epsbkb \epsakb  \nonumber\\
 &+&  3 \xd{12} \epsbkb \epsaqb+\xd{12} \epsbqa \epsakb-
           4 \xd{13} \epsbkb \epsakb\nonumber\\
 &-&  2 \xd{13} \epsbkb \epsaqb-2 \xd{13} \epsbqa \epsakb-\xd{13} \epsaepsb s
\nonumber\\
 &+&  2 \xd{22} \epsbkb \epsaqb-\xd{22} \epsaepsb t-2 \xd{23} \epsbqa \epsakb
\nonumber\\
 &-&  2 \xd{23} \epsbqa \epsaqb-\xd{23} \epsaepsb t+6 \xd{24} \epsbkb \epsakb
\nonumber\\
 &+&  3 \xd{24} \epsbkb \epsaqb+\xd{24} \epsbqa \epsakb
            -6 \xd{25} \epsbkb \epsakb\nonumber\\
 &-&  2 \xd{25} \epsbkb \epsaqb-2 \xd{25} \epsbqa \epsakb-\xd{25} \epsaepsb s
\nonumber\\
 &-&  4 \xd{26} \epsbkb \epsaqb+4 \xd{26} \epsbqa \epsakb
            +2 \xd{26} \epsbqa \epsaqb\nonumber\\
 &+& \xd{26} \epsaepsb \(s+2 t\)-\xd{32} \epsaepsb q_2^2+
      \xd{33} \epsaepsb q_1^2\nonumber\\
 &+&  2 \xd{34} \epsbkb \epsakb-2 \xd{35} \epsbkb \epsakb\nonumber\\
 &+&    \xd{36} \epsaepsb \(q_2^2-t\)-
      2 \xd{37} \epsbqa \epsakb\nonumber\\
 &+&   2 \xd{36} \epsbkb \epsaqb +\xd{37} \epsaepsb \(q_2^2-s-t\)\nonumber\\
 &+&  2 \xd{38} \epsbqa \epsaqb+\xd{38} \epsaepsb \(q_1^2+2 q_2^2-s\)
\nonumber\\
 &-&  2 \xd{39} \epsbqa \epsaqb-\xd{39} \epsaepsb \(2 q_1^2+q_2^2-s\)
\nonumber\\
 &-&  2 \xd{310} \epsbkb \epsaqb+2 \xd{310} \epsbqa \epsakb\nonumber\\
 &-&  \xd{310} \epsaepsb\(2 q_2^2-s-2 t\)+4 \xd{312} \epsaepsb\nonumber\\
 &-&  4 \xd{313} \epsaepsb\Big] \;,
\ea
\ba
 c_b &=& -2 \Big\{ \lq \xd{36}+\xd{37}-2 \xd{310}\rq \(q_2^2-t\)
\nonumber\\
 &+&  \xd{38} \(q_1^2+2 q_2^2\) - \xd{39} \(2 q_1^2+q_2^2\) \Big\}-
      2 \Big[\xd0\nonumber\\
 &+&  \xd{11}+\xd{12}-2 \xd{13}+\xd{24}-\xd{25}\nonumber\\
 &+&\xd{26}-\xd{37}-\xd{38}+\xd{39}+\xd{310}\Big] s\nonumber\\
 &+& 2 \Big\{\lq \xd{22}+\xd{23}-2 \xd{26}\rq t - 2 \xd{27} \nonumber\\
 &+& \xd{32} q_2^2 -\xd{33} q_1^2- 6 \(\xd{312}-\xd{313}\)\Big\}
\nonumber\\
 &-& \frac1t \lq T_b(q_1^2,t)+T_b(q_2^2,t)+\tilde B_0(t)
                   -5+\frac{\pi^2}{3} \; \rq\,,
\ea
with
\bq
T_b(q^2,t) = \frac{1}{t-q^2}\lg
2q^2\lq\tilde B_0(t)-\tilde B_0(q^2)\rq
+t\tilde B_0(t)-q^2\tilde B_0(q^2)\rg-2q^2\tilde C_0(q^2,t) \,.
\eq
For the crossed function $\tilde{\cal M}(q_2,q_1)$, the same expressions 
as above apply, with the obvious interchange $q_1\leftrightarrow q_2$,
$\eps_1\leftrightarrow \eps_2$, and
$t\,\to\, u$.

The finite part of the $D_0$ function is defined by
\bq
\label{eq:d0tilde}
\tilde D_0(k_2,q_2,q_1) = \frac{1}{2st}\lq
 \ln^2\frac{q_1^2q_2^2}{t^2}
+4\;{\rm Li}_2\(1-\frac{t}{q_1^2}\)
+4\;{\rm Li}_2\(1-\frac{t}{q_2^2}\)-\frac{\pi^2}{3} \rq\; .
\eq
This expression is well defined when all invariants, $q_1^2$, $q_2^2$ and $t$,
are space-like. In our application, we always have one space-like and
one time-like weak boson, i.e., exactly one of the two quotients $t/q_i^2$
is positive. In the other quotient simply replace the time-like invariant by
$t\,\to \, t+i0^+$ or $q_i^2\,\to\, q_i^2+i0^+$, as in 
Eqs.~(\ref{eq:b0tilde}) and~(\ref{eq:c0tilde}).

The remaining finite $\tilde D_{ij}$ functions are obtained from the above 
expressions for the $\tilde B_0$, $\tilde C_0$, and $\tilde D_0$ functions
with the usual Passarino-Veltman recursion relations given in 
Ref.~\cite{Passarino:1978jh},
adapted to the Bjorken-Drell metric, $q_i^2>0$ for a time-like momentum $q_i$.
In these recursion relations we need the additional finite $\tilde B_0$ and 
$\tilde C_0$ functions
\ba
\tilde B_0(0) &=& 0\;,\\
\tilde C_0(k_2,q_1+q_2) &=& \tilde C_0(s,0,0) = \frac{1}{s}\frac{\pi^2}{6}\;,
\ea
while
\bq
\tilde C_0(q_1,q_2) = C_0(q_1^2,q_2^2,s) 
\eq
is the infrared- and ultraviolet-finite $C_0$ function for massless
internal propagators but with nonzero invariants $q_1^2$, $q_2^2$ and $s$.

%%%%%%%%%%%%%%%%%%%%%%%%%%%%%%%%%%%%%%%%%%%%%%%%%%%%%%%%%%%%%%%%%%%%%%%%%%%%% 


\begin{thebibliography}{99}

\bibitem{ATLAS}
ATLAS Collaboration, ATLAS TDR,
%ATLAS Detector and Physics Performance Technical Design Report,
Report No.\ CERN/LHCC/99-15 (1999);
E.~Richter-Was and M.~Sapinski,
Acta Phys.\ Pol.\ B {\bf 30}, 1001 (1999);
%%CITATION = APPOA,B30,1001;%%
B.~P.~Kersevan and E.~Richter-Was,
Eur.\ Phys.\ J.\ C {\bf 25}, 379 (2002)
[arXiv:hep-ph/0203148].
%%CITATION = HEP-PH 0203148;%%

\bibitem{CMS}
G.~L.~Bayatian {\it et al.}, CMS Technical Proposal,
Report No.\ CERN/LHCC/94-38x (1994);
%D. Denegri, %Prospects for Higgs (SM and MSSM) searches at LHC,
%talk in the Circle Line Tour Series, Fermilab, October 1999,
%(http://www-theory.fnal.gov/CircleLine/DanielBG.html);
R.~Kinnunen and D.~Denegri, %Expected SM/SUSY Higgs observability in CMS,
CMS Note No.\ 1997/057;
R. Kinnunen and A. Nikitenko,
%Study of $H_{SUSY}\to \tau\tau\to l^{\pm}+h^{\mp} + E_t^{miss}$ in CMS,
Report No.\ CMS TN/97-106;
R.~Kinnunen and D.~Denegri,
%The $H_{SUSY}\to \tau\tau\to h^{\pm}+h^{\mp}+X$
%channel, its advantages and potential instrumental drawbacks,
arXiv:hep-ph/9907291;
%%CITATION = HEP-PH 9907291;%%
V.~Drollinger, T.~M\"uller and D.~Denegri,
arXiv:hep-ph/0111312.
%%CITATION = HEP-PH 0111312;%%

\bibitem{Zeppenfeld:2000td}
D.~Zeppenfeld, R.~Kinnunen, A.~Nikitenko and E.~Richter-Was,
%``Measuring Higgs boson couplings at the LHC,''
Phys.\ Rev.\ D {\bf 62}, 013009 (2000)
[arXiv:hep-ph/0002036];
%%CITATION = HEP-PH 0002036;%%
D.~Zeppenfeld,
%``Higgs couplings at the LHC,''
in {\it Proc. of the APS/DPF/DPB Summer Study on the Future of
Particle Physics, Snowmass, 2001} edited by N.~Graf,
eConf {\bf C010630}, p.\ 123 (2001)
[arXiv:hep-ph/0203123];
%%CITATION = HEP-PH 0203123;%%
A.~Belyaev and L.~Reina,
%``p p $\to$ t anti-t H, H $\to$ tau+ tau-: Toward a model independent
%determination of the Higgs boson couplings at the LHC,''
JHEP {\bf 0208}, 041 (2002)
[arXiv:hep-ph/0205270].
%%CITATION = HEP-PH 0205270;%%

%\cite{Rainwater:1999gg}
\bibitem{Rainwater:1999gg}
D.~L.~Rainwater,
%``Intermediate-mass Higgs searches in weak boson fusion,''
arXiv:hep-ph/9908378.
%%CITATION = HEP-PH 9908378;%%

%\cite{Campbell:2002tg}
\bibitem{Campbell:2002tg}
J.~Campbell and R.~K.~Ellis,
 %``Next-to-leading order corrections to W + 2jet and Z + 2jet production  at
%hadron colliders,''
Phys.\ Rev.\ D {\bf 65}, 113007 (2002)
[arXiv:hep-ph/0202176];
%%CITATION = HEP-PH 0202176;%%
J.~Campbell, R.~K.~Ellis and D.~Rainwater,
 %``Next-to-leading order QCD predictions for W + 2jet and Z + 2jet production
%at the CERN LHC,''
Phys.\ Rev.\ D {\bf 68}, 094021 (2003)
[arXiv:hep-ph/0308195].
%%CITATION = HEP-PH 0308195;%%

%\cite{Chehime:1992ub}
\bibitem{Chehime:1992ub}
H.~Chehime and D.~Zeppenfeld,
%``Single W and Z boson production as a probe for rapidity gaps at the SSC,''
Phys.\ Rev.\ D {\bf 47}, 3898 (1993).
%%CITATION = PHRVA,D47,3898;%%

%\cite{Rainwater:1996ud}
\bibitem{Rainwater:1996ud}
D.~Rainwater, R.~Szalapski and D.~Zeppenfeld,
%``Probing color-singlet exchange in Z + 2-jet events at the LHC,''
Phys.\ Rev.\ D {\bf 54}, 6680 (1996)
[arXiv:hep-ph/9605444].
%%CITATION = HEP-PH 9605444;%%

%\cite{Khoze:2002fa}
\bibitem{Khoze:2002fa}
V.~A.~Khoze, M.~G.~Ryskin, W.~J.~Stirling and P.~H.~Williams,
%``A Z monitor to calibrate Higgs production via vector boson fusion with  rapidity gaps at the LHC,
Eur.\ Phys.\ J.\ C {\bf 26}, 429 (2003)
[arXiv:hep-ph/0207365].
%%CITATION = HEP-PH 0207365;%%

\bibitem{Baur:1993fv}
U.~Baur and D.~Zeppenfeld,
%``Measuring three vector boson couplings in q q $\to$ q q W at the SSC,''
arXiv:hep-ph/9309227.
%%CITATION = HEP-PH 9309227;%%

%\cite{Figy:2003nv}
\bibitem{wbfhtautau}
D.~Rainwater, D.~Zeppenfeld and K.~Hagiwara,
%``Searching for H $\to$ tau tau in weak boson fusion at the LHC,''
Phys.\ Rev.\ D {\bf 59}, 014037 (1999) [arXiv:hep-ph/9808468];
%%CITATION = HEP-PH 9808468;
T.~Plehn, D.~Rainwater and D.~Zeppenfeld,
%``A method for identifying H --> tau tau --> e+- mu-+ missing p(T)
%at the CERN LHC,''
Phys.\ Rev.\ D {\bf 61}, 093005 (2000) [arXiv:hep-ph/9911385];
%[hep-ph/9911385].
S.~Asai et al., Report No.\ ATL-PHYS-2003-005.

\bibitem{wbfhtoww}
D.~Rainwater and D.~Zeppenfeld,
%``Observing $H \to W^{(*)}W^{(*)} \to e^\pm \mu^\mp /\!\!\!{p}_T$ in weak boson fusion with dual fo
Phys.\ Rev.\ D {\bf 60}, 113004 (1999)
[Erratum-ibid.\ D {\bf 61}, 099901 (2000)]
[arXiv:hep-ph/9906218];
%%CITATION = HEP-PH 9906218;%%
N.~Kauer, T.~Plehn, D.~Rainwater and D.~Zeppenfeld,
%``H $\to$ W W as the discovery mode for a light Higgs boson,''
Phys.\ Lett.\ B {\bf 503}, 113 (2001)
[arXiv:hep-ph/0012351];
%%CITATION = HEP-PH 0012351;%%
C.~M.~Buttar, R.~S.~Harper and K.~Jakobs,
Report No.\ ATL-PHYS-2002-033;
K.~Cranmer et al.,
Report No.\ ATL-PHYS-2003-002 and Report No.\ ATL-PHYS-2003-007;
S.~Asai et al.,
Report No.\ ATL-PHYS-2003-005.

\bibitem{Eboli:2000ze}
O.~J.~Eboli and D.~Zeppenfeld,
%``Observing an invisible Higgs boson,''
Phys.\ Lett.\ B {\bf 495}, 147 (2000)
[arXiv:hep-ph/0009158];
%%CITATION = HEP-PH 0009158;%%
B.~Di Girolamo, A.~Nikitenko, L.~Neukermans, K.~Mazumdar and D.~Zeppenfeld,
%``Experimental observation of an invisible Higgs boson at LHC,''
in arXiv:hep-ph/0203056.

%\cite{Baur:1993fv}
\bibitem{Charlton:2001am}
D.~G.~Charlton,
%``Experimental tests of the standard model,''
arXiv:hep-ex/0110086.
%%CITATION = HEP-EX 0110086;%%
The LEP Electroweak Working Group: {\tt http://lepewwg.web.cern.ch/LEPEWWG}.

\bibitem{Han:1992hr}
T.~Han, G.~Valencia and S.~Willenbrock,
%``Structure function approach to vector boson scattering in p p collisions,''
Phys.\ Rev.\ Lett.\  {\bf 69}, 3274 (1992)
[arXiv:hep-ph/9206246].
%%CITATION = HEP-PH 9206246;%%


%\cite{Figy:2003nv}
\bibitem{Figy:2003nv}
T.~Figy, C.~Oleari and D.~Zeppenfeld,
 %``Next-to-leading order jet distributions for Higgs boson production via
%weak-boson fusion,''
Phys.\ Rev.\ D {\bf 68}, 073005 (2003)
[arXiv:hep-ph/0306109].
%%CITATION = HEP-PH 0306109;%%

\bibitem{Boudjema:1996qg}
F.~Boudjema {\it et al.},
%``Standard Model Processes at LEP-2,''
arXiv:hep-ph/9601224.
%%CITATION = HEP-PH 9601224;%%

%\cite{Charlton:2001am}
\bibitem{madgraph}
T.~Stelzer and W.~F.~Long,
%``Automatic generation of tree level helicity amplitudes,''
Comput.\ Phys.\ Commun.\  {\bf 81}, 357 (1994)
[arXiv:hep-ph/9401258];
%%CITATION = HEP-PH 9401258;%%
F.~Maltoni and T.~Stelzer,
%``MadEvent: Automatic event generation with MadGraph,''
JHEP {\bf 0302}, 027 (2003)
[arXiv:hep-ph/0208156].
%%CITATION = HEP-PH 0208156;%%


\bibitem{HZ}
K.~Hagiwara and D.~Zeppenfeld,
%``Helicity Amplitudes For Heavy Lepton Production In E+ E- Annihilation,''
Nucl.\ Phys.\ {\bf B274}, 1 (1986);
%%CITATION = NUPHA,B274,1;%%
K.~Hagiwara and D.~Zeppenfeld,
%``Amplitudes For Multiparton Processes Involving A Current ...,''
Nucl.\ Phys.\  {\bf B313}, 560 (1989).
%%CITATION = NUPHA,B313,560;%%

\bibitem{CS}
S.~Catani and M.~H.~Seymour,
%``A general algorithm for calculating jet cross sections in NLO QCD,''
Nucl.\ Phys.\  {\bf B485}, 291 (1997)
[Erratum-ibid.\  {\bf B510}, 503 (1997)]
[arXiv:hep-ph/9605323].
%%CITATION = HEP-PH 9605323;%%

\bibitem{DR_citation}
Warren Siegel, Phys.\ Lett.\ B {\bf 84}, 193 (1979);
%%CITATION = PHLTA,B84,193;%%
Warren Siegel, Phys.\ Lett.\ B {\bf 94}, 37 (1980).
%%CITATION = PHLTA,B94,37;%%

\bibitem{ilyin}
V.~Ilyin, private communication.


\bibitem{vegas}
G.~P. Lepage, J. Comput. Phys. {\bf 27},  192  (1978).

%\cite{Argyres:1995ym}
\bibitem{Argyres:1995ym}
See, e.g., E.~N.~Argyres {\it et al.},
%``Stable calculations for unstable particles: Restoring gauge invariance,''
Phys.\ Lett.\ B {\bf 358}, 339 (1995)
[arXiv:hep-ph/9507216].
%%CITATION = HEP-PH 9507216;%%

%\cite{Baur:1991pp}
\bibitem{Baur:1991pp}
U.~Baur, J.~A.~Vermaseren and D.~Zeppenfeld,
%``Electroweak vector boson production in high-energy e p collisions,''
Nucl.\ Phys.\  {\bf B375}, 3 (1992).
%%CITATION = NUPHA,B375,3;%%

%\cite{Denner:1999gp}
\bibitem{Denner:1999gp}
A.~Denner, S.~Dittmaier, M.~Roth and D.~Wackeroth,
%``Predictions for all processes e+ e- $\to$ 4fermions + gamma,''
Nucl.\ Phys.\  {\bf B560}, 33 (1999)
[arXiv:hep-ph/9904472].
%%CITATION = HEP-PH 9904472;%%

\bibitem{lopez}
See, e.g., G.~Lopez Castro, J.L.M.~Lucio and J.~Pestieau,
Mod.~Phys.~Lett. {\bf A6}, 3679 (1991);
M.~Nowakowski and A.~Pilaftsis, Z.~Phys. {\bf C60}, 121 (1993);
U.~Baur and D.~Zeppenfeld,
%``Finite width effects and gauge invariance in radiative W production...''
Phys.\ Rev.\ Lett.\  {\bf 75}, 1002 (1995)
[arXiv:hep-ph/9503344],
%%CITATION = HEP-PH 9503344;%%
and references therein.

%\cite{Hagiwara:1986vm}
\bibitem{Hagiwara:1986vm}
K.~Hagiwara, R.~D.~Peccei, D.~Zeppenfeld and K.~Hikasa,
%``Probing The Weak Boson Sector In E+ E- $\to$ W+ W-,''
Nucl.\ Phys.\  {\bf B282}, 253 (1987).
%%CITATION = NUPHA,B282,253;%%

\bibitem{kniehl}
K.~Hagiwara, D.~Zeppenfeld and S.~Komamiya,
%``Excited Lepton Production At Lep And Hera,''
Z.\ Phys.\ C {\bf 29}, 115 (1985);
%%CITATION = ZEPYA,C29,115;%%
B.~A.~Kniehl,
%``Elastic E P Scattering And The Weizsacker-Williams Approximation,''
Phys.\ Lett.\ B {\bf 254}, 267 (1991).
%%CITATION = PHLTA,B254,267;%%

\bibitem{cteq6}
J.~Pumplin, D.~R.~Stump, J.~Huston, H.~L.~Lai, P.~Nadolsky and W.~K.~Tung,
%``New generation of parton distributions with uncertainties ...,''
JHEP {\bf 0207}, 012 (2002)
[arXiv:hep-ph/0201195].
%%CITATION = HEP-PH 0201195;%%


\bibitem{PDG}
K.~Hagiwara {\it et al.}  [Particle Data Group Collaboration],
%``Review Of Particle Physics,''
Phys.\ Rev.\ D {\bf 66}, 010001 (2002).
%%CITATION = PHRVA,D66,010001;%%


\bibitem{kToriginal}
S.~Catani, Yu.~L.~Dokshitzer and B.~R.~Webber,
                Phys.\ Lett.\ B  {\bf 285} 291 (1992);
S.~Catani, Yu.~L. Dokshitzer, M.~H.~Seymour and B.~R.~Webber,
                Nucl.\ Phys.\  {\bf B406} 187 (1993);
S.~D.~Ellis and D.~E.~Soper, Phys.\ Rev.\ D
{\bf 48} 3160 (1993).

\bibitem{kTrunII}
G.~C.~Blazey {\it et al.},
%``Run II jet physics,''
arXiv:hep-ex/0005012.
%%CITATION = HEP-EX 0005012;%%

\bibitem{mrst}
A.~D.~Martin, R.~G.~Roberts, W.~J.~Stirling and R.~S.~Thorne,
%``Uncertainties of predictions from parton distributions. I: Experimental
%errors,''
Eur.\ Phys.\ J.\ C {\bf 28}, 455 (2003)
[arXiv:hep-ph/0211080];
%%CITATION = HEP-PH 0211080;%%
A.~D.~Martin, R.~G.~Roberts, W.~J.~Stirling and R.~S.~Thorne,
%``Uncertainties of predictions from parton distributions. II: Theoretical
%errors,''
arXiv:hep-ph/0308087.
%%CITATION = HEP-PH 0308087;%%

\bibitem{Passarino:1978jh}
G.~Passarino and M.~J.~Veltman,
%``One Loop Corrections For E+ E- Annihilation Into Mu+ Mu- In The Weinberg
%Model,''
Nucl.\ Phys.\ {\bf B160}, 151 (1979).
%%CITATION = NUPHA,B160,151;%%

\end{thebibliography}
\end{document}